%% file: main.tex
\documentclass[runningheads,table,orivec]{llncs}
\usepackage[paperheight=235mm, paperwidth=155mm,textwidth=12.2cm,textheight=19.3cm]{geometry}

\usepackage{styles/ProCount}
\usepackage{styles/tensors}
\usepackage{styles/smallsec}
\usepackage{amsmath}

\usepackage{amsthm}
\usepackage{graphicx}
\usepackage{tikz}
\usepackage{mathtools}
\usepackage{amssymb}
\usepackage{xcolor}
\usepackage[T1]{fontenc}
\usepackage{MnSymbol}
\usepackage{stmaryrd}
\usepackage[]{algorithmicx}
\usepackage{algpseudocode}
\usepackage{todonotes}
\usepackage{bbding}
\usepackage{float}
\usepackage{subfigure}
\usepackage{todonotes}

\title{Dynamic Boolean Synthesis with Zero-suppressed Decision Diagrams} 

\author{Yi Lin \inst{1}\orcidID{0000-0001-8443-2246} \and 
Moshe Y. Vardi \Envelope \inst{1}\orcidID{0000-0002-0661-5773}}
\authorrunning{Y. Lin et al.}
\institute{Rice University, Houston TX 77005, USA \\
\email{yl182@rice.edu,vardi@cs.rice.edu}}

\begin{document}
\maketitle              

\begin{abstract}
Motivated by functional synthesis in sequential circuit construction and quantified boolean formulas (QBF), boolean synthesis serves as one of the core problems in Formal Methods. Recent advances show that decision diagrams (DD) are particularly competitive in symbolic approaches for boolean synthesis, among which zero-suppressed decision diagram (ZDD) is a relatively new algorithmic approach, but is complementary to the industrial portfolio, where binary decision diagrams (BDDs) are more often applied. We propose a new dynamic-programming ZDD-based framework in the context of boolean synthesis, show solutions to theoretical challenges, develop a tool, and investigate the experimental performance. We also propose an idea of magic number that functions as the upper bound of planning-phase time and tree width, showing how to interpret the exploration-exploitation dilemma in planning-execution synthesis framework. The algorithm we propose shows its strengths in general, gives inspiration for future needs to determine industrial magic numbers, and justifies that the framework we propose is an appropriate addition to the industrial synthesis solvers portfolio.

\keywords{Boolean synthesis \and Binary decision diagram \and Zero-suppressed binary decision diagram \and Exploration-exploitation \and Dynamic programming.}
\end{abstract}
%
%

\input{Introduction}

\input{Preliminaries}
\input{Realizability}
\input{Synthesis}
\input{Experiments}
\input{Conclusion}

\newpage
\bibliographystyle{splncs04}
\bibliography{list}

\appendix
\input{Appendix}

\vfill

\end{document}

%% file: Introduction.tex
\section{Introduction}\label{sec:introduction}

In computing scenarios, a description of boolean-value relations between known and unknown variables does not guarantee a constructive representation for the outputs. The problem of taking in a declarative constraint and generating solutions for the unknowns in term of the knowns, called \emph{boolean synthesis}~\cite{hachtel2007logic}, has been investigated in multiple fields. A well-designed synthesis framework can contribute to the construction of missing components in an electronic circuit as long as the desired behavior is abstracted into a boolean formula~\cite{akshay2021boolean}. In certificate construction of quantified boolean formulas (QBFs), synthesis algorithms can function as an inspiration or get inspired by QBF solving approaches~\cite{Lucas_Functional_Synthesis,RabeS16_QBF}. When a sequential circuit is to be constructed responding to time-sensitive environment inputs, a temporal synthesis algorithm solves for the system outputs that satisfy the temporal specification~\cite{ZhuTLPV17,ZhuTLPV17a}, where boolean synthesis is a building block. 

Many previous works~\cite{CAV16,Kukula_Circuits} and recent advances in boolean-synthesis research have proposed valuable algorithms, including those applying machine-learning techniques~\cite{golia2020manthan,manthan2}, knowledge-compilation mechanisms~\cite{akshay2019knowledge}, or QBF solving~\cite{DBLP:conf/sat/RabeS16}. We, in this work, focus on the pursuit of a \emph{symbolic, dynamic-programming-based, scalable and space-efficient} solution, on top of the exploration in previous relevant works~\cite{ZDDpaper,DPSynthpaper}. 

A popular approach in boolean reasoning is the use of decision diagrams, such as binary decision diagrams (BDDs) \cite{bdd}; such approached are often referred to as \emph{symbolic}.
Many variant of BDDs have been studied, such as algebraic decision diagrams (ADDs)~\cite{bahar1997algebric}, zero-suppressed decision diagrams (ZDDs)~\cite{MinatoZDD}, and sentential decision diagrams (SDDs)~\cite{DBLP:conf/aaai/ChoiD13} also investigated in synthesis-related topics~\cite{dudek2020addmc,DBLP:conf/ijcai/GiacomoFLVX022,ZDDpaper}. Several recent works explored the use of symbolic methods in boolean synthesis, such as~\cite{CAV16,FMCAD17,ZDDpaper}.
These symbolic approaches for representing boolean functions have the common advantage that not only the solutions for unknown variables are synthesized but also the \emph{realizability set} (i.e., the set of input values for which there exist output values satisfying the specification). This advantage is imperative functionality for synthesis algorithms, as in \emph{modular circuit design} for the purpose of confirming the consistency of outputs in independently constructed modules~\cite{hachtel2007logic}, and as in temporal synthesis \cite{ZhuTLPV17,ZhuTLPV17a} for iterative union of realizability sets in winning sets construction. Mature implementation tools~\cite{CUDD_cudd} also exist for converting boolean formulas into these decision diagrams, which is another advantage of using symbolic approaches. Moreover, \emph{partially realizable} specifications with an input domain that is neither \emph{fully} nor \emph{nullary} realizable are also solvable when BDD-like representations are used.

The BDD-based approaches, however, often face the challenge of exponential blowups~\cite{FMCAD17,ZDDpaper}. This impact makes the approach infeasible when the numbers of variables and clauses scale up. Hence, we take the alternative decision-diagram structure of ZDD, as proposed for symbolic boolean synthesis and shown to be complementary and scalably promising in the portfolio with BDDs~\cite{ZDDpaper}, as well as bringing in better performance in QBF solving~\cite{QMRes_DBLP:conf/cp/PanV04}. The ZDD structure preserves the structural advantages of BDD in synthesis solving as above, while it adds compactness to the representation with smaller size than BDD for identical formulas, providing an approach that is complementary to the BDD-based approach.

But ZDDs are also accompanied by unique challenges compared to BDDs. Its structural design represents clauses instead of truth assignments, enabling compact representation of CNF formulas. But this makes it more difficult to test whether the realizability set is empty or not \cite{ZDDpaper}. Also, quantifier elimination over a ZDD may involve an exponential blow-up.

As boolean formulas can be recursively simplified into smaller or simpler problems before final synthesis, also in order to reduce space impact, we adopt the dynamic-programming (DP) approach in~\cite{dudek2020addmc,dudek2020dpmc} to recursively break down input formulas into smaller components. A similar investigation was reported for BDD-based boolean synthesis~\cite{DPSynthpaper}, but adopting this approach to ZDDs poses new algorithmic challenges. Building on the tree-decomposition approach of ~\cite{dudek2020addmc,dudek2020dpmc}, we propose a novel  synthesis algorithm with both realizability checking and witness constructions. As substitutions of variables in ZDD usually easily change the \emph{form} (e.g. CNF and DNF) and modify the interpretation of a ZDD representation, we have to overcome this issue here in the dynamic-programming framework. 

Earlier works~\cite{FMCAD17,DPSynthpaper} compared compare symbolic dynamic-programming-based tools against non-symbolic non-synthesis tools, including CegarSkolem~\cite{cegarskolem}, BFSS~\cite{BFSS} and Manthan~\cite{Manthan}, and were shown to be a significant addition to the boolean-synthesis solver portfolio. Here, our experiments focus on comparison with other symbolic approaches, specifically, algorithms a BDD-based DP-based tool~\cite{DPSynthpaper} and a non-DP ZDD-based tool~\cite{ZDDpaper}. 

The dynamic-programming approach consists of two phases. In the \emph{planning} phase, a good tree decomposition of the input formulas is identified. Then, in the \emph{execution phase}, the tree is used to guide the dynamic-programming evaluation. Our experiments reveal that we are facing here the classical \emph{exploration-exploitation dilemma}~\cite{explorationexploitation}, as getting a tree decomposition of low width, which enables efficient evaluation, is itself a computationally challenging task. We address this dilemma here by discovering, experimentally, a \emph{magic number}, to guide the planning phase. We demonstrate the advantages of this approach but also discuss its limits.  We conclude that a dynamic-programming ZDD-based tool is an important addition to the portfolio of solvers for boolean synthesis.


%% file: Preliminaries.tex
\section{Preliminaries}\label{sec:preliminaries}


Consider a \emph{boolean formula} $\varphi(X)$.  Let $AP(\varphi)$ denote the set of atomic propositions in a formula ${\varphi}$. We assume some order on its atomic propositions $AP(\varphi)$. Let $X=AP(\varphi)=(x_1, \ldots, x_m)$, then $\varphi$ represents a boolean function $f : \mathbb{B}^m \rightarrow \mathbb{B}$.
Substitution in $\varphi(X)$ of a formula $\theta(X)$ in place of a variable $x_i$ is denoted by $\varphi[x_i \mapsto \theta]$ and defined by $\varphi[x_i \mapsto \theta](X) = \varphi(x_1, \ldots, x_{i-1}, \theta(X), x_{i+1}, \ldots, x_m)$. Two boolean formulas $\varphi$ and $\varphi'$ are logically equivalent if they represent the same boolean function (and therefore have the same set of satisfying assignments).




 A boolean formula has literals and boolean operators. A \emph{literal} is either a boolean variable or the negation of a variable. A \emph{clause} is a disjunction of literals, and a \emph{cube} is a conjunction of literals. Boolean formulas in the form of a conjunction of clauses are said to be in \emph{Conjunctive Normal Form} (CNF), and a boolean formula in the form of a disjunction of cubes is said to be in \emph{Disjunctive Normal Form} (DNF). We abuse notation and consider a CNF formula as a set of clauses, and a DNF formula as a set of cubes. Both clauses and cubes can be viewed as sets of literals. A clause $c_1$ \emph{subsumes} a clause $c_2$ if $c_1\subseteq c_2$, and thus $c_1$ logically implies $c_2$. Hence, if a CNF formula contains $c_1$, then $c_2$ can be removed while preserving logical equivalence. We can thus assume that our CNF formulas are \emph{subsumption free}; that is, no clause is subsumed by another. Note that the empty clause, which is considered unsatisfiable, subsumes all clauses. 
 
We now define the major problem we solve in this paper. 
\begin{definition}[Boolean Synthesis Problem] \label{def:boolean-synthesis}
Given a boolean formula $\varphi(\vec{x}, \vec{y})$ with $m + n$ boolean variables, partitioned into $m$ \emph{input variables} $\vec{x} = (x_1, \ldots, x_m)$ and $n$ \emph{output variables} $\vec{y} = (y_1, \ldots, y_n)$, construct:
\begin{enumerate}
    \item The set $R \subseteq \mathbb{B}^m$, called the \emph{realizability set}, of all assignments $\vec{a} \in \mathbb{B}^m$ to $\vec{x}$ for which there exists an assignment $\vec{b} \in \mathbb{B}^n$ to $\vec{y}$ such that $\varphi(\vec{a}, \vec{b}) = 1$. 
    \item A function $g : \mathbb{B}^m \rightarrow \mathbb{B}^n$ such that $\varphi(\vec{a}, g(\vec{a})) = 1$ for all $\vec{a} \in R$. This is called a \emph{witness function}. 
\end{enumerate}
\end{definition}
In practice, arbitrary Boolean formulas can be converted to equi-realizable CNF formulas with a linear blowup using Tseytin encoding, quantifying existentially over Tseytin variables \cite{prestwich2009cnf}. The witnesses for the equi-realizable formula can then be used for the original formula. Thus, we focus in this paper on input formulas in CNF. In addition to CNF and DNF representations, other symbolic representations have been proposed for Boolean-function manipulation, among which one of the most popular is that of Binary Decision Diagrams~\cite{bryant1986graph}.

\paragraph{Binary Decision Diagrams.}
A (Reduced Ordered) Binary Decision Diagram (BDD) is a directed acyclic
graph. Internal nodes of the BDD represent boolean variables. Internal nodes have two children, representing the assignments 0 and 1.
Root-to-terminal paths on the BDD correspond to truth assignments, leading either to a terminal node 1 if satisfying or a terminal node 0 if unsatisfying. We assume that all BDDs are \emph{ordered}, meaning that variables are ordered in the same way along every path, and \emph{reduced}, meaning that superfluous nodes are removed and identical subgraphs are merged. Given these two conditions, BDDs are a \emph{canonical} representation, meaning that two BDDs with the same variable order that represent the same function are identical. The variable order used can have a major impact on the BDD's size, and two BDDs representing the same function but with different orders can have an exponential difference in size.


\noindent \paragraph{ Zero-Suppressed Decision Diagrams (ZDDs).}
BDDs represent the set of satisfying assignment of a propositional formula. Another approach is to represent the set of clauses of a CNF formua \cite{minato1993zero}. A propositional variable $p$ has three truth values in a clause $c$: no occurrence, positive occurrence, or negative occurrence, which can be represented by two Boolean variables, $x_p$ and $x_{\neg p}$. If $p$ does not  occur in $c$, then both $x_P$ and $x_{\neg p}$ should be false, and $x_P$ and $x_{\neg p}$ should never be both true. Thus, a CNF formula with $n$ propositional variables can be represented by a BDD over $2n$ boolean variables. 

In many application clauses are \emph{sparse}; that is, most clauses contain at most $k$ literals, for some $k<<n$. Since ``no occurrence'' is the most common status of a propositional variable in sparse clause, represented by the assignment $00$ to the two corresponding Boolean variables, it makes sense to make $0$ the default truth value. \emph{Zero-suppressed decision diagrams} (ZDDs) are a variant of BDDs, but with $0$ as the default value \cite{mishchenko2001introduction}.  ZDDs are similar to BDDs but use a different reduction rule: while BDDs remove nodes where both edges point to the same child, in ZDDs \emph{also} a node is replaced with its negative child if its positive edge points to the terminal node 0. Thus, a CNF formula, $\varphi$, viewed as a set of clauses, can be represented compactly as a ZDD  $\llbracket \varphi\rrbracket$ \cite{ZRes_DBLP:conf/cade/ChatalicS00}. While every root-terminal path in the BDD $B_\varphi$ of $\varphi$ represents a satisfying assignment of $\varphi$, every every root-terminal path in $\llbracket \varphi\rrbracket$ a clause of $\varphi$.  


Note that there are two ``extreme'' ZDDs of special interest:
(1) if $\varphi$ is the empty formula satisfied by all assignments, then $\llbracket \varphi \rrbracket$ has a single terminal $0$, and (2) if $\varphi=\{\emptyset\}$, it is satisfied by no assignment, so $\llbracket \varphi \rrbracket$, as a ZDD, is the single terminal $1$, meaning that it contains the unsatisfiable empty clase. This is completely unique to the ZDD structure and opposite to BDD interpretations, where the BDD 1 represents all truth assignments and the BDD 0s represents the empty set of truth assignments. 
We describe operations on ZDDs in Section~\ref{subsec:zdd operations}
and Section.~\ref{section:synth}. 

Next, we describe the realizability and synthesis problems.  

\begin{definition}[Realizability Set]\label{def: realiza set}
Given a CNF formula $\varphi(X,Y)$ over \emph{input} and {\em output} variables $X$ and $Y$, respectively, the \emph{realizability set} of $\varphi$ \emph{with respect to $Y$}, denoted $R_\varphi(Y) \subseteq 2^X$, is defined to be the set of assignments $\sigma \in 2^X$ for which there exists an assignment $\tau \in 2^Y$ such that $\varphi(\sigma \bigcup \tau) = 1$. When $\varphi$ is clear from context, we simply denote the realizability set $R_{\varphi}$.
\end{definition}


\begin{definition}[Full, Partial and Nullary Realizability]\label{def: full partial null real}
    Let $\varphi(X,Y)$ be a CNF formula with $X$ and $Y$ as input and output variables. We say that $\varphi$ is \emph{fully realizable} if $R_\varphi = 2^X$. We say that $\varphi$ is \emph{partially realizable} if $R_\varphi \neq \emptyset$. Finally, we say that $\varphi$ is \emph{nullary realizable} if $R_\varphi = \emptyset$.
\end{definition}

Given the condition that the formula is at least partially realizable, 
the boolean synthesis problem asks for a set of \emph{witnesses} for the output variables generated as functions of the input variables, such that the formula is satisfied. This sub-problem of constructing the witness functions, is usually referred to as synthesis. 

\begin{definition}[Witnesses in Boolean Synthesis Problem]\label{def: witness construction}
    Let $\varphi(X,Y)$ denote a fully or partially realizable boolean formula  with input variables in $X=\{x_1,\ldots,x_m\}$ and output variables $Y =\{y_1,\ldots,y_n\}$, and let $R_{\varphi}(X) \neq \emptyset$ be its realizability set. A sequence  $g_1(X), \ldots, g_n(X)$ of boolean functions is a sequence of \emph{witness functions} for the $Y$ variables in $\varphi(X,Y)$ if for every assignment ${x} \in R_{\varphi}(X)$, we have that
    $\varphi[X \mapsto {x}][y_1 \mapsto g_1({x})]\ldots[y_n \mapsto g_n({x})]$ holds.
\end{definition}

The procedure of algorithmically constructing the witnesses is then defined to be boolean synthesis.

\begin{definition}[Synthesis]\label{def: synthesis}
     Given a partially or fully realizable CNF formula $\varphi(X,Y)$ with input and output variables $X$ and $Y$, the \emph{synthesis problem} asks to algorithmically construct a set of \emph{witness functions} for the $Y$ variables in terms of the $X$ variables.     
\end{definition}

For a CNF formula $\varphi=c_1 \wedge \ldots \wedge c_n$, the realizability set is  expressed by the quantified formula ${\Sigma}_Y (c_1 \wedge \ldots \wedge c_n)$. Thus, the realizability set can be computed using two operations: conjunction and existential quantification. The \emph{dynamic-programming} approach aims at scheduling these operations as to minimize the cost of intermediate operations. To that end we introduce a graph-theoretic structure that guides in scheduling conjunction and quantification operations.


%% file: Realizability.tex
\section{ZDD Realizability Algorithm}\label{sec:realizability}

In order to describe our dynamic-programming framework, we introduce the concepts of project-join trees and its graded variant.

\paragraph{Project-join tree of a CNF formula.}
\cite{dudek2020dpmc}
Given a CNF boolean formula $\varphi(X, Y)$, a \emph{project-join tree} for the formula is defined as a tuple $\T = (T, r, \gamma, \pi)$, where:
\begin{enumerate}
    \item $T$ is a tree with a set $\V{T}$ of vertices, a set $\Lv{T} \subseteq \V{T}$ of leaves and a root $r \in \V{T}$.
    \item $\gamma : \Lv{T} \to \varphi$ is a bijection that maps the leaves of $T$ to the clauses of $\varphi$.
    \item $\pi : \V{T} \setminus \Lv{T} \to 2^X$ is a function that labels internal nodes with variable sets, where the labels $\{\pi(n) \mid n \in \V{T} \setminus \Lv{T}\}$ form a partition of $X$.
    \item If a clause $c \in \varphi$ contains a variable $x$ that belongs to the label $\pi(n)$ of an internal node $n \in \V{T} \setminus \Lv{T}$, then the associated leaf node $\gamma^{-1}(c)$ must descend from $n$.
\end{enumerate} 


\paragraph{Graded project-join tree of a CNF formula.}\cite{Procount}
If the variables in $\varphi$ are partitioned into $(X,Y)$, and if there are quantifiers applied to the formula, then it would be more convenient to perform operations on the trees, if variables in the $X$-graded are placed on top of the $Y$-graded variables in the construction of the tree. A generalization of project-join trees based on this idea is that \emph{graded} project-join tree.

A project-join tree $\T = (T, r, \gamma, \pi)$ of a CNF formula $\varphi(X,Y)$ over variables $X\bigcup Y$, where $X\cap Y=\emptyset$, is \emph{($X$, $Y$)-graded} if there exist \emph{grades} $\I_X, \I_Y \subseteq \V{T}$ that partition the internal nodes $\V{T} \setminus \Lv{T}$, such that:
    \begin{enumerate}
        \item The grade of a node is consistent with its label; that is, if $n_X \in \I_X$ then $\pi(n_X) \subseteq X$, and if $n_Y \in \I_Y$ then $\pi(n_Y) \subseteq Y$.
        \item $\I_X$ is above $\I_Y$; that is, if $n_X \in \I_X$ and $n_Y \in \I_Y$, then $n_X$ is not a descendant of $n_Y$ in $T$.
    \end{enumerate}
\noindent \textbf{Notation for graded project-join trees}:
            \begin{itemize}
                \item $C(n), D(n), P(n)$: the children set, descendants set, and parent node of a given node $n$, respectively.
                \item $\T_n$: the subtree of tree $\T$ rooted at node $n \in \V{T}$, including all nodes and arcs below $n$.
                \item $\texttt{YTreeRoots}(\T)$: the highest-level (closest to the root) $Y$-graded internal nodes, formally $\texttt{YTreeRoots}(\T) = \{n \in \I_Y \mid (n = r) \lor (P(n) \in \I_X)\}$. The root is in this set if and only if the partition $X$ is empty. We call this set $\texttt{YTreeRoots}$ because they match the set of roots of the subtrees consisting only of nodes from $Y$-partition.    
                \item Conversely, $\texttt{XTreeLeaves}(\T) = \{n \in \I_X \mid C(n) \subseteq \I_Y)\}$ is the set of lowest-level $X$-graded internal nodes.
            \end{itemize}



\setlength{\floatsep}{7pt plus 2pt minus 2pt}
\setlength{\textfloatsep}{7pt plus 2pt minus 2pt}
\setlength{\intextsep}{7pt plus 2pt minus 2pt}






\subsection{ZDD Operations and Notations}\label{subsec:zdd operations}
Before introducing the algorithm to compute and check realizability properties for the symbolic ZDD representation, we specify a list of operations that are uniquely defined for the ZDD encoding of boolean formulas. 
\begin{definition}
\textbf{Projection}: 
               Given CNF formula ${\varphi}$ and a variable set $X$, we use $\Sigma_X\varphi$ to denote a quantified boolean formula (QBF)         expressing a boolean function that maps a truth assignment $\tau \in \mathcal{B}^{{AP}_{\varphi} - X}$ to $({\Sigma}_X \varphi)(\tau)$, 
                        defined as                         
                        $\bigvee_{\mathcal{X}\in {\mathcal{B}^X}} {(\varphi(\tau) \lor \varphi(\tau \cup \mathcal{X})})$, where $\mathcal{B} = \{0,1\}$, i.e., it is true when there is an assignment $\mathcal{X}$ to $X$ that is part of a satisfying assignment of ${\varphi}$.  We call this ${\Sigma}_X{\varphi} $ the \textbf{projection} on CNF formula ${\varphi}$ with respect to variables in $X$. 
                        \end{definition}

We now show how to compute projection for ZDD representation of formulas.
\begin{definition}[ZDD Operations \cite{minato1993zero,Clause_Distribution_and_Subsum_Free_DBLP:conf/ictai/ChatalicS00,mishchenko2001introduction,KNUTH_10.5555/1593023})]\label{def: ZDD operations}

        \begin{itemize}
                
        \item \textbf{Selection}: 
        The following operations select clauses in a formula $\varphi$ with respect to a variable $y$:
        $\llbracket{\varphi_y^+ }\rrbracket\equiv Subset1(z,y)$,  $\llbracket{\varphi_y^- }\rrbracket \equiv Subset1(z,\neg y)$, and $\llbracket{ \varphi_y'}\rrbracket \equiv Subset0(Subset0(z,y), \neg y)$. 
        Similarly, 
        if $\varphi$ is represented by ZDD $Z$, then $Z_y^+$ denotes the ZDD representing $\varphi_y^+$, and $Z_y^-$ denotes the ZDD representing $\varphi_y^-$.

        
        \item \textbf{Subsumption-free union}: Given two ZDDs $z_1$ and $z_2$, representing two sets of clauses, subsumption-free union $\bigcup_{sf}$ takes the union of these two sets of clauses, after removing the subsumed redundant clauses. The result of this operation is denoted by $(z_1 \bigcup_{sf} z_2).$

        \item \textbf{Clause Distribution}: This operation is equivalent to the disjunction of $z_1$ and $z_2$ encoding two sets of clauses, denoted by $(z_1 \times_{cd} z_2)$.
    
        
        

    \end{itemize}
\end{definition}

\begin{lemma}
\label{lemma: resolution}{\bf Symbolic Resolution}
For a CNF formula ${\varphi}$ with a \emph{single} boolean variable $y$ that occurs in this formula, the ZDD of
${\Sigma}_y{\varphi} $ is equivalent to a representation of $(( \llbracket{ \varphi^+_y }\rrbracket \times_{cd} \llbracket{ \varphi^-_y }\rrbracket) \bigcup_{sf} \llbracket{ \varphi'_y }\rrbracket)$. 

\end{lemma}

To compute the projection with respect to a set $X$ of variables, we apply Lemma~\ref{lemma: resolution} and iterate over all variables of $X$.

\begin{definition}[ZDD-Valuations of Nodes in Graded Project Join Tree]\label{def: ZDD pair valuations}
    We define a pair of mutually related valuations, labeling nodes by ZDDs.
    
    Let $\T = (T, r, \gamma, \pi)$ has the $\I_X$ and $\I_Y$. The \emph{post-valuation} of a node $n\in T$ is defined as
    $$\texttt{BV}_{\texttt{post}}(\T, n)=
        \begin{cases}
            
		\llbracket{\gamma(n) }\rrbracket, \\
                 &\text{if $n \in \Lv{\T_n}$}\\
            {\Sigma}_{\pi(n)}  (\texttt{BV}_{\texttt{pre}}(\T, n)), \\
                 &\text{if $n \in (\I_X \bigcup \I_Y)$}
		 \end{cases}$$\\
    The \emph{pre-valuation} of $n$ is defined as
    $$\texttt{BV}_{\texttt{pre}}(\T, n)=
        \begin{cases}
		\llbracket{\gamma(n) }\rrbracket, 
                \hspace{1cm}
                &\text{if $n \in \Lv{\T_n}$}, \\
            ( \texttt{BV}_{\texttt{post}}(n'_1) \bigcup_{sf} \ldots \bigcup_{sf} \texttt{BV}_{\texttt{post}}(n'_m) ), \\
                \hspace{1cm} \text{where } C(n)=\{n'_1, \ldots, n'_m\},
                &\text{if $n \in (\I_X \bigcup \I_Y)$} \\
		 \end{cases}$$
\end{definition}
That is, to compute the ZDD $\texttt{BV}_{\texttt{post}}(\T, n)$, 
we use symbolic resolution.
Since a ZDD represents a set of clauses, we carry out conjunction in the computation of the ZDD $\texttt{BV}_{\texttt{pre}}(\T, n)$
via the subsumption-free union $\bigcup_{sf}$ for obtaining $\texttt{BV}_{\texttt{pre}}(\T, n)$.

\paragraph{Notations on sub-trees:} Denote the sub-tree of graded project-join tree $\T$ rooted at a node $n \in \mathcal{V}(T)$ by $\T_n$.
We also use the notation $\gamma \mid_n$ and $\pi \mid_n$. The domain of $\gamma$ is the set of leaves $\Lv{T}$, which is a superset of $\Lv{T_n} = \{ m \mid m \in \Lv{T} \land m \in D(n) \}$. So we define $\gamma \mid_n$ to be the projected function of $\gamma $ on the smaller domain $\Lv{T_n}$. Similarly, as the domain is $\pi$ is $\mathcal{V}(T) \setminus \Lv{T}$, which is a superset of $\mathcal{V}(T_n) \setminus \Lv{T_n}$, we define $\pi \mid_n$ to be the projected function of $\pi $ on the smaller domain $\mathcal{V}(T_n) \setminus \Lv{T_n}$.

\subsection{Realizability Set}\label{subsec:realizability set}
\noindent Pre-valuation for a node in a graded tree corresponds to the conjunction of children post-valuations, while a post-valuation for the node matches the result of quantifying variables in its label. The step of conjoining ensures that quantifications can always be performed, and the leaf nodes encoding clauses containing variables $\pi(n)$ in the label of a node $n$ are always the descendants of that node. Moreover, all nodes in $\I_Y$ occur below all nodes in $\I_X$.

\begin{algorithm}[h]
\caption{$\texttt{ComputeValuations}(\T_n)$}\label{alg: compute valuations}
    \DontPrintSemicolon
    \SetKwFunction{this}{ComputeValuations}
    \SetKwInOut{Function}{Operation}
    \SetKwInOut{Parameter}{Notation}

    \KwIn{a sub-tree $\T_n = (T\mid_n, n, \gamma\mid_n, \pi \mid_n)$ rooted at a node $n \in \I_Y$ of an ($(X,Y)$-graded) project-join tree $\T = (T, r, \gamma, \pi)$ of a CNF $\varphi$}
    \KwOut{$\texttt{BV}_{\texttt{post}}(\T_n, n)$ and $\texttt{BV}_{\texttt{pre}}(\T_n, n)$ for $n$ with respect to $\T_n$, with possibly early determination of nullary realizability for $\varphi$}

    \uIf {\upshape $n \in \Lv{T_n}$}{
        $\texttt{pre-BV}(\T_n,n),\texttt{post-BV}(\T_n,n) \gets \gamma\mid_n(n)$\tcp{base case: is a leaf}\label{line leaf}

        \Return{}\;

    } \uElse {
        $\texttt{post-BV}(\T_n,n), \texttt{pre-BV}(\T_n,n) \gets \{\}$\tcp{set both to tautology}
        \For{$m \in C(n)$}{
            \tcp{recursively compute the valuations for children of $n$}
            \this($\T_m = (T_m, m, \gamma \mid_m, \pi \mid_m)$)\;\label{line recursive call}
            
            \tcp{if UNSAT in the sub-tree under $n$}
            \uIf {$\texttt{post-BV}(\T_m,m) == \{\emptyset\}$}{
                
                \Return{nullary realizable}\tcp{early determination}\label{line return 1}
            } \Else {
                \tcp{update pre-valuation of $r$}
                $\texttt{pre-BV}(\T_n,n) \gets ( \texttt{pre-BV}(\T_n,n) \bigcup_{sf} \texttt{post-BV}(\T_m, m) )$

                \tcp{if pre-valuation of a subset of the clauses is UNSAT}
                \tcp{if is already checked since union does not bring more sets containing an emptyset}
            }   
        }
        \tcp{obtain pre-valuation of $n$ with respect to $\T_n$}
        $\texttt{post-BV}(\T_n,n) \gets \texttt{pre-BV}(\T_n,n)$\;
        \tcp{projecting out variables on children labels}
        $\texttt{post-BV}(\T_n,n) \gets \Sigma_{\pi\mid_n(n)}(\texttt{post-BV}(\T_n,n))$\;
                    
        \If {$\texttt{post-BV}(\T_n, n) == \{\emptyset\}$}{
                        \Return{nullary realizable}\;\label{line return 3}
                    }
        \Return{}\tcp{return computed valuations if realizability is NOT null }\label{line return}
    } 
\end{algorithm}

We therefore apply in Algorithm.\ref{alg: compute valuations} the operations of Definition.~\ref{def: ZDD pair valuations} on $\I_Y$ nodes of the tree, before computing realizability set in the next algorithm and apply this definition again in checking realizability properties.

In a graded project-join tree, we can see from the perspective of $I_X, I_Y$ partition that the parents of uppermost $I_Y$ nodes are the lowermost $I_X$ nodes, along the paths from the root to the leaves. Hence we denote by two specific notations for convenience in this paper: 
$\texttt{YTreeRoots}(\T)$ and $\texttt{XTreeLeaves}(\T)$, as defined at the beginning of Section~\ref{sec:realizability}.

To compute the ZDD valuations, we first apply $\texttt{ComputeValuations}$ to the highest-level $Y$-graded internal nodes of an ($(X,Y)$-graded) project-join tree $\T = (T, r, \gamma, \pi)$ of a CNF formula $\varphi$. That is, for each of $n \in \texttt{YTreeRoots}(\T)$, we apply $\texttt{ComputeValuations}(\T_{n})$ to the subtree $\T_n = (T\mid_n, n, \gamma\mid_n, \pi \mid_n)$ rooted at $n$. We compute the valuations $\texttt{BV}_{\texttt{post}}(\T_n, n)$ and $\texttt{BV}_{\texttt{pre}}(\T_n, n)$ for $n$ with respect to $\T_n$, with possibly early determination of nullary realizability for $\varphi$.

\begin{theorem}\label{thm: alg computeValuation correctness}
Let $\T = (T, r, \gamma, \pi)$ be a graded project-join tree of a CNF formula $\varphi$. Then applying Algorithm~\ref{alg: compute valuations} on a node $n \in \texttt{YTreeRoots}(\T)$ returns $\texttt{BV}_{\texttt{post}}(\T_n, n)$ and $\texttt{BV}_{\texttt{pre}}(\T_n, n)$, or correctly early determines nullary realizability.
\end{theorem}

Note that since we label variables by literal pairs in our ZDD encoding, in practice we project out this pair of literals together. The algorithm above shows the preprocessing traversing through leaves to the root and computing pairs of valuations for every node. The next algorithm depicts the process of getting realizability set.

\begin{algorithm}[h]
\caption{$getRSet(\T)$}\label{alg: compute zr}
    \DontPrintSemicolon
    \SetKwFunction{this}{getRSet}
    \SetKwFunction{AlgOne}{ComputeValuations}
    \SetKwInOut{Function}{Operation}
    \SetKwInOut{Parameter}{Notation}
    \SetKwFunction{XTreeLeaves}{XTreeLeaves}
    \SetKwFunction{YTreeRoots}{YTreeRoots}
    \SetKwFunction{pre}{\texttt{pre-BV}}
    \SetKwFunction{post}{\texttt{post-BV}}
    
    \KwIn{$\T = (T, r, \gamma, \pi)$: an ($(X,Y)$-graded) project-join tree representing a CNF formula}
    \KwOut{the ZDD for union of post-valuations of $\texttt{YTreeRoots}(\T)$}

    $\llbracket R \rrbracket \gets \{\}$\tcp{initialized to be $\{\}$}
    $\T_{\texttt{new}} \gets \T$\;
    \For {$n \in \texttt{YTreeRoots}(\T)$}
    {
        $\T_n = (T\mid_n, n, \gamma\mid_n, \pi
\mid_n)$\;
        \If{$\AlgOne(\T_n)$ returns nullary realizable\label{line alg 1}}{
            \Return{$\{\emptyset\}$}\tcp{without need for further synthesis}
        }
        
        $\llbracket R \rrbracket \gets (\llbracket R \rrbracket \bigcup_{sf} \post(\T_n, n))$\label{line union}\;
                        
                    $\mathcal{V}(T_{\texttt{new}}) \gets \mathcal{V}(T_{\texttt{new}})\setminus D(n)$\tcp{remove $n$ descendants}\label{line update 1}
        
                    $\Lv{T_{\texttt{new}}} \gets \Lv{T_{\texttt{new}}}\bigcup\{n\}$\tcp{add $n$ to leaves of $\T_{\texttt{new}}$}\label{line update 2}
    }
            
            \Return{$\llbracket R \rrbracket, \T_{\texttt{new}}$}
\end{algorithm}

\begin{algorithm}[h]
\caption{$\texttt{checkPartial}(\T)$}\label{alg: realize zr}
    \DontPrintSemicolon
    \SetKwFunction{this}{checkPartial}
    \SetKwFunction{AlgOne}{ComputeValuations}
    \SetKwFunction{AlgTwo}{getRSet}
    \SetKwInOut{Function}{Operation}
    \SetKwInOut{Parameter}{Notation}
    \SetKwFunction{XTreeLeaves}{XTreeLeaves}
    \SetKwFunction{YTreeRoots}{YTreeRoots}
    \SetKwFunction{pre}{\texttt{pre-BV}}
    \SetKwFunction{post}{\texttt{post-BV}}
    
    \KwIn{$\T = (T, r, \gamma, \pi)$: an ($(X,Y)$-graded) project-join tree representing a CNF formula $\varphi$}
    \KwOut{Determination of nullary, full, or partially realizability. Returns realizability set ZDD $\llbracket R \rrbracket$ in partially realizable case.}

    $\llbracket R \rrbracket, \T_{\texttt{new}} \gets \AlgTwo(\T)$\label{line {} alg 3}\;
    
    \uIf{$\llbracket R \rrbracket == \{\}$}{
        \Return{fully realizable}\label{line return 1 2.1}
    } \Else {
                \tcp{Break Point: from this line on, either proceed as below, OR materialze clause set and send it to a SAT solver}\label{line break point}\label{line: SAT solver}
                \tcp{The following operation performs post-valuation on the $X$ tree, with inputs labelled on $\I_X$ quantified}
                
                Update $\post(\T_{\texttt{new}}, r)$ by $\AlgOne(\T_{\texttt{new}})$\;
                \uIf{$\post(\T_{\texttt{new}}, r) == \{\emptyset\}$}{
                    \Return{nullary realizable}\label{line return 1 3rd}
                }\Else{
                    %
                    \Return{partially realizable}

                }
    }
\end{algorithm}         
    


\begin{theorem}\label{thm: alg RealizeZR correctness}
Let $\T = (T, r, \gamma, \pi)$ be a graded project-join tree of CNF $\varphi$. Assume that Algorithm~\ref{alg: compute valuations} returns the correct ZDDs for pairs of valuations by \texttt{ComputeValuations}. 
Then:
$\llbracket R \rrbracket$ returned by Algorithm~\ref{alg: compute zr} is a ZDD that represents the subsumption-free union of $\texttt{post-BV}(\T, n) )$ for all $n \in \texttt{YTreeRoots}(\T)$, which is the correct realizability set of $\varphi$.
\end{theorem}

\begin{theorem}\label{thm: alg compute zdd of R correctness}
    Algorithm~\ref{alg: compute zr} and Algorithm~\ref{alg: realize zr} returns the correct identification among full, partial and nullary realizability status, and, in partially-realizable cases, the final value of $\llbracket R \rrbracket$ is the correct realizability set of $\varphi$.
    
\end{theorem}

%% file: Synthesis.tex
\section{Synthesis of Witnesses}\label{section:synth}



\subsection{Monolithic Techniques for ZDD-based Synthesis}\label{subsec: monolithic synth}

The core component of a boolean-synthesis solution, as defined in Definition \ref{def:boolean-synthesis}, is to construct a multidimensional witness function $g: \mathbb{B}^m \rightarrow \mathbb{B}^n$ for the set of output variables $\vec{y}$ as a function of the inputs $\vec{x}$. In this section, we describe a witness-construction algorithm, with a correctness proof, using the graded project-join trees and intermediate ZDD representations, assuming a partially or fully realizable domain as identified in the phase of realizability checking.

We first describe a monolithic technique in witness constructions for boolean synthesis~\cite{ZDDpaper}.
We want to solve for multiple output variables $y_1, \ldots, y_m$ in a formula $\varphi$, which is encoded by ZDD $Z$. First, starting from $Z$, existential quantification is performed with respect to the output variables inside-out, by applying the projection operation defined in Definition~\ref{def: ZDD operations}. Formally, given a ZDD $Z(X, y_1, \ldots, y_m)$ for $\varphi$, we project $y_1, \ldots, y_m$ in order as follows:

\vspace{-1mm}
\allowdisplaybreaks
\begin{align*}
    &Z_0(X, y_1, \ldots, y_m) = Z(X, y_1, \ldots, y_m), \\
     & Z_1(X, y_2, \ldots, y_m)=(\Sigma_{y_1})Z_0,\\
     &\ldots\\
     &Z_{m-1}(X, y_m)=(\Sigma_{y_{m-1}})Z_{m-2},\\
     &Z_m(X)=(\Sigma_{y_m})Z_{m-1}. 
\end{align*}

We start with the witness function for $y_m$ in $Z_{m-1}(X, y_m)$. When there is a single output variable to solve, previous work~\cite{ZDDpaper} demonstrated two CNF witnesses, based on Lemma.~\ref{lemma: resolution}.
\begin{lemma}\label{lemma: witness one y}
Let $X$ be a set of input variables and $y$ a single output variable in a Boolean CNF formula $\varphi$ with $R_\phi(Y) \neq \emptyset$. Then $\varphi_y^-$ and $\neg \varphi_y^+$ are both witnesses for $y$ in $\varphi$.
\end{lemma}
After the witness function is obtained by Lemma.~\ref{lemma: witness one y}, substitution of $g_m$ into $Z_{m-2}(X,y_{m-1}, y_m)$ generates a ZDD representing the formula $(\Sigma_{y_{m-2}})\ldots(\Sigma_{y_1}) \\Z[y_m \mapsto g_m]$, where the only variable to solve is $y_{m-1}$. We then get the witness functions for $y_m, \ldots, y_1$ by applying substitutions iteratively~\cite{ZDDpaper}.

For the process to be applicable iteratively, we need to ensure that the substitutions preserve conjunctive normal form (CNF).
Thus, we substitute $g_m$, which is in CNF, for positive occurrences of $y_m$ and we substitute $(\neg g_m+)$, which is in DNF, for negative occurrences of $y_m$. 
Since explicit conversion between CNF to DNF involves an exponential blowup, we perform the conversion symbolically, using Knuth's \texttt{Cross} operations. The \texttt{Cross} operation is defined to be a set operation that, first, eliminates the sets that are subsets of other sets, from a set of sets, and, next, constructs a set of new sets that each have at least one common element with each of the set after the first step. In \cite{ZDDpaper} we identify and prove this operation to be interpretable as a way to construct the DNFs that, when interpreted on their ZDDs, are equivalent encodings of the CNFs.

With the iterative construction of $Z_m$, as the output variables are projected out in order, witnesses are constructed in the reverse order, and so in our case $y_m, \ldots, y_1$ are substituted into $Z_{i}$, from which we then construct the next witnesses. Formally, $g_m$ is computed from:
$Z'_{m-1}(X,y_m) = Z_{m-1}(X,y_m),$  via $g_{m}={Z'_{m-1}}_{y_m}^-;$

$\ldots$

$g_{i}$ is computed from: $Z'_{i-1}(X,y_i)=Z_{i-1}(X,y_1,\ldots,y_i)[y_{i+1} \mapsto g_{i+1}]$
$\ldots[y_{m}\mapsto g_{m}],$ via $g_{i}={Z'_{i-1}}_{y_i}^-.$

The following lemma is based on \cite{CAV16}.
\begin{lemma}\label{lemma: monolithic correct}
If $R_\varphi(Y)\not=\emptyset$, then
the $g_i$'s above are witness functions for $Y$ in $\phi(X,Y)$.
\end{lemma}

\subsection{Algorithm for ZDD-based Synthesis on Graded Project-join Trees decompositions}



Our framework works for both partial and fully realizable cases. 
We first compute realizability sets going \emph{bottom up} the graded project-join tree for $\phi$ as in Section 3.
We now show that the witness functions for the $Y$ variables can then be constructed by iterated substitution \emph{top-down}. In other words, we first compute the witnesses for the variables in the labels of internal nodes at higher levels, and then propagate those down toward the leaves. Note that in the graded-project-join-trees framework for projected model counting, which inspired our approach here, the trees are processed only in a bottom-up fashion \cite{Procount}. In contrast, here we need to combine bottom-up and top-down processing.

The following Lemma~\ref{lemma: independence} provides the essential relation among witness functions for different variables.

\begin{lemma}\label{lemma: independence}
Let $m$ and $n$ be two distinct internal nodes in a graded project-join tree $\T$ such that $n$ is not a descendant of $m$. Then no variable in $\pi(m)$ appears in $\texttt{BV}_{\texttt{pre}}(\T, n)$.
\end{lemma}
It follows from the lemma that the witness for a variable $y_i \in \pi(n)$ computed from $\texttt{BV}_{\texttt{pre}}(\T, n)$, depends only on witnesses of output variables $y_j \in \pi(m)$ where $n$ is a descendant of $m$.

\paragraph{Synthesizing Inner-node Variables.}
We now adapt the monolithic approach to the dynamic-programming framework. Denote the label for node $n$ to be $\pi(n)$. Then to formalize the sub-problem of solving for variables $y_1, \ldots, y_m \in \pi(n)$, we are given a ZDD $\texttt{BV}_\texttt{pre}(\T, n)=Z(\vec{x}, y_1, \ldots, y_m, \vec{y}')$, where $\vec{y}'$ are the variables not labeled on $n$ but on its ancestors up to $\texttt{YTreeRoots}(\T)$. We treat $\vec{y}'$ in the same way as $\vec{x}$, whose reason is explained above. Then:

\allowdisplaybreaks
\begin{align*}
    &Z = Z(X, y_1, \ldots, y_m, \vec{y}'), \\
     &(\Sigma_{y_1})Z = Z_1(X, y_2, \ldots, y_m, \vec{y}'), \\
     &\ldots\\
     &(\Sigma_{y_{m-1}})\ldots(\Sigma_{y_1})Z = Z_{m-1}(X, y_m, \vec{y}')\\
     &(\Sigma_{y_m})(\Sigma_{y_{m-1}})\ldots(\Sigma_{y_1})Z = Z_m(X, \vec{y}'). 
\end{align*}

By a similar set of projections and substitutions as in Subsection~\ref{subsec: monolithic synth}, we obtain a set of temporary witnesses, in which $\vec{y}'$ are to be substituted later in the algorithm by witnesses to those variables labeled on ancestors of $n$, for an arbitrary node $n \not\in \texttt{YTreeRoots}(\T)$.

\paragraph{Synthesizing on Inter-node Variables.}
All $y \in \pi(n)$ can be synthesized from $\texttt{BV}_\texttt{pre}(\T, n)$ as a monolithic specification. For the variables labeled on a node $n \in \texttt{YTreeRoots}(\T)$, they occur in clauses represented by leaves that are descendants of $n$, but the variables labeled on lower-layer descendants have only occurrences on narrower scope of leaves. Thus, we can take the witnesses constructed for variables on $n$ into $\texttt{BV}_\texttt{pre}(\T, n')$ and solve for $y \in \pi(n')$ for children and descendants $n'$ of $n$, and then substitute later using the witnesses for these output variables labeled on descendants.

\vspace{0.1cm}

\begin{algorithm}[H]
\caption{$DPZynth(\T)$}\label{alg: synth}
    \DontPrintSemicolon
    \SetKwFunction{this}{DPZynth}
    \SetKwFunction{AlgOne}{ComputeValuations}
    \SetKwInOut{Function}{Operation}
    \SetKwInOut{Parameter}{Notation}
    \SetKwFunction{XTreeLeaves}{XTreeLeaves}
    \SetKwFunction{YTreeRoots}{YTreeRoots}
    \SetKwFunction{pre}{\texttt{pre-BV}}
    \SetKwFunction{post}{\texttt{post-BV}}
    
    \KwIn{pre-valuations of $n \in \texttt{YTreeRoots}(\T)$, where $\T = (T, r, \gamma, \pi)$ is an ($(X,Y)$-graded) project-join tree of a CNF formula $\phi(\vec{x},\vec{y})$ that has partially or fully realizable domain}
    \KwOut{the ZDDs for witnesses $g_{i,CNF}$ for all $y_i \in \vec{y}$ }
    
    $V \gets \{\}$\;
    \For {$n \in \texttt{YTreeRoots}(\T)$}
    {
        Add $n$ to the end of $V$\;
        \For {$n' \in D(n)$}
        {
            \If {$n' \not\in \Lv{\T}$}
            {
                Add $n'$ to the end of $V$\;
            }
        }
    }\label{line: got V}
    \For {$n \in V$}
    {\label{line for loop start}
        existentially project out $y \in \pi(n)$ from $\texttt{BV}_\texttt{pre}(\T, n)$, and get $Z_1\ldots Z_{m-1}$ until $Z_m$ with only one output variable $y_m \in \pi(n)$ left\;
        $g_{m,CNF} \gets {Z_m}_{y_m}^-$,
        $g_{m, DNF} \gets \texttt{Cross}(g_{m,CNF})$\;
        
        Substitute $y_m$ and $\neg y_m$ by $g_{m,CNF}$ and $g_{m, DNF}$ into $Z_{m-1}$\;
        $g_{m-1,CNF}\gets {Z_{m-1}}_{y_{m-1}}^-$,
        $g_{m-1, DNF}\gets \texttt{Cross}(g_{m-1,CNF})$\;
        
        Iteratively, until obtain $g_{1,CNF}\gets {Z_{1}}_{y_1}^-$,
        $g_{1, DNF}\gets \texttt{Cross}(g_{1,CNF})$\;\label{line on-node labeled vars}
        
        \For {$y' \in A(n)$ where $A(n)$ stands for the set of ancestors of $n$ up to nodes in $ \texttt{YTreeRoots}(\T)$}
        {
            $g_{i,CNF} \gets g_{i,CNF}$ with literal $y'$ mapped to $g_{y',CNF}$ and literal $\neg y'$ mapped to $g_{y',DNF}$\;\label{line A(n) substitution}
        }

    }\label{line for loop ends}
    \Return{$g_{i,CNF},\forall y_i \in \vec{y}$}
\end{algorithm}

\vspace{0.1cm}

\paragraph{Algorithm.}
Based on these insights, we present in Algorithm~\ref{alg: synth} our dynamic-programming synthesis algorithm for producing the witnesses $g_y(X)$ for each output variable $y$, represented by a ZDD. Starting from nodes $n \in \texttt{YTreeRoots}(\T)$, for variables $y \in \pi(n)$, we obtain the witness for $y$ from the pairs of valuations for descendants of all $n \in \texttt{YTreeRoots}(\T)$ which cover the constraints relevant to all $y$ in the original CNF specification, as explained above. Other variables are treated in the same way as input variables, as they are labeled on descendants of $\texttt{YTreeRoots}(\T)$.

We obtain CNF witnesses $g_{i, CNF}$ for all $y_i$ labeled on $n \in \texttt{YTreeRoots}(\T)$ by Lemma.~\ref{lemma: witness one y} and have the corresponding DNF witnesses $g_{i, DNF}$ by Knuth's \texttt{Cross} operation interpreted in \cite{ZDDpaper}, as Section~\ref{subsec: monolithic synth}. The computed paired witnesses preserve an overall CNF and equivalence after substitutions to positive and negative literals respectively. Applying pre-valuations for $n' \in C(n), n \in \texttt{YTreeRoots}(\T)$, where the pre-valuation formulas are the utilized as the original ZDDs as in the monolithic approach, the variables labeled on the children $n'$ of $n \in \texttt{YTreeRoots}(\T)$ are synthesized.

The operations are performed until we get to the nodes that are direct parents of leaf nodes that represent the partitioned formula, i.e., clauses. Then all output variables are synthesized and substituted into previous witnesses constructed for variables labeled on ancestor nodes, in order to obtain a full set of solutions that do not contain occurrences of $\vec{x}$ variables. We put this algorithm in Algorithm~\ref{alg: synth}.

\begin{theorem}\label{thm: synth alg}
Given a CNF formula $\phi$ and a graded project-join tree $\T = (T, r, \gamma, \pi)$ of $\phi$. If the input domain is partially or fully realizable, Algorithm~\ref{alg: synth} returns a set of witness functions for outputs $\vec{y}$ in $\phi$.
\end{theorem}

%% file: Experiments.tex
\section{Experimental Evaluation}\label{sec:experiments}

From previous works~\cite{FriedTV16,FMCAD17,ZDDpaper,DPSynthpaper}, we know the following. First, decision-diagram-based techniques are competitive with other techniques that are based on search and learning. Second, among symbolic approaches, BDDs and ZDDs are complementary. Third, in the context of BDD-based synthesis, dynamic-programming techniques offer an advantage. Based on the theoretical framework of the previous sections, we developed a tool, \emph{DPZynth}. Our goal in this section is to examine whether the dynamic-programming approach offers an advantage for ZDD-based synthesis by comparing DPZynth to the monolithic ZDD-based tool ZSynth \cite{ZDDpaper}. We also compare the performance of DPZynth to that of the dynamic-programming BDD-based tool DPSynth \cite{DPSynthpaper} to study the relative advantage of BDDs and ZDDs in the dynamic-programming framework.

The challenge in evaluating the advantage of the dynamic programming approach is that it consists of two phases. In the \emph{planning phase}, we use tree decomposition to obtain the graded project-join tree, while in the \emph{execution phase} we utilize this tree for realizability testing and synthesis. Spending more time on planning may decrease execution time, so there is a trade-off between planning time and execution time. 
This trade-off is essentially the well-known \emph{exploration-exploitation dilemma}~\cite{explorationexploitation}. 

We resolved this trade-off here by identifying the \emph{magic number} of $200$: an upper bound of $200$s second for the planning phase, while searching for a tree of tree width below $200$. This magic number is obtained by running experiments on our benchmark suite, as explained below, in the sprit of \emph{programming by optimization} \cite{hoos2012programming}. (This means that this magic number is workload-dependent.) 


\emph{Experimental Settings and Methodology}: Using the CUDD BDD library~\cite{CUDD_cudd}, the FlowCutter tree-decomposition tool~\cite{flowcutter}, and the MCS variable ordering~\cite{tabajara2019partitioning} as default, we compared DPZynth to the monolithic ZDD-based tool ZSynth~\cite{ZDDpaper}, and to a graded-tree-based BDD boolean synthesizer DPSynth~\cite{DPSynthpaper}. The 275 boolean synthesis benchmarks are taken from the latest purely forall-exists data set of $\Pi^P_2$ CNFs from QBFEVAL~\cite{qbfeval} in 2016 to 2022, without additional selection criteria, with additional parametric integer factorization and subtraction families generated by \cite{akshay2021boolean,supratik2}. The experiments were run on double-blind university
clusters, assigning simultaneously to a mix of hardware with 16-96 cores with 32-256 GB RAM, with each solver-benchmark combination on a single core. Each benchmark is given an end-tp-end maximal time of two hours.

\subsection{Magic Numbers}

We are interested in the following two questions:
\begin{itemize}
    \item[(1)] What is an appropriate timeout limit for the tree-decomposition phase?
    \item[(2)] What tree-width bound should we set when looking for tree decompositions?
\end{itemize}

We define \emph{planning overhead} to be the time cost in looking for an appropriate tree decomposition. In practice, the overhead incurred by tree decomposition is usually the cost on time or space. We use planning time here, as space performance is usually consistent with time in boolean synthesis. 

We define a \emph{winning tree} to be a graded project-join tree found by the tree decomposition tool that brings an ultimate shorter \emph{end-to-end} realizability-checking time than that of the monolithic tool ZSynth. \emph{End-to-end} time here includes preprocessing time on format-conversion of benchmarks, compilation from input formula to ZDD representation, tree-decomposition time, and realizability-checking time.

\begin{figure}[h]
\centering
\begin{subfigure}[b]
\centering
\includegraphics[width=\textwidth]{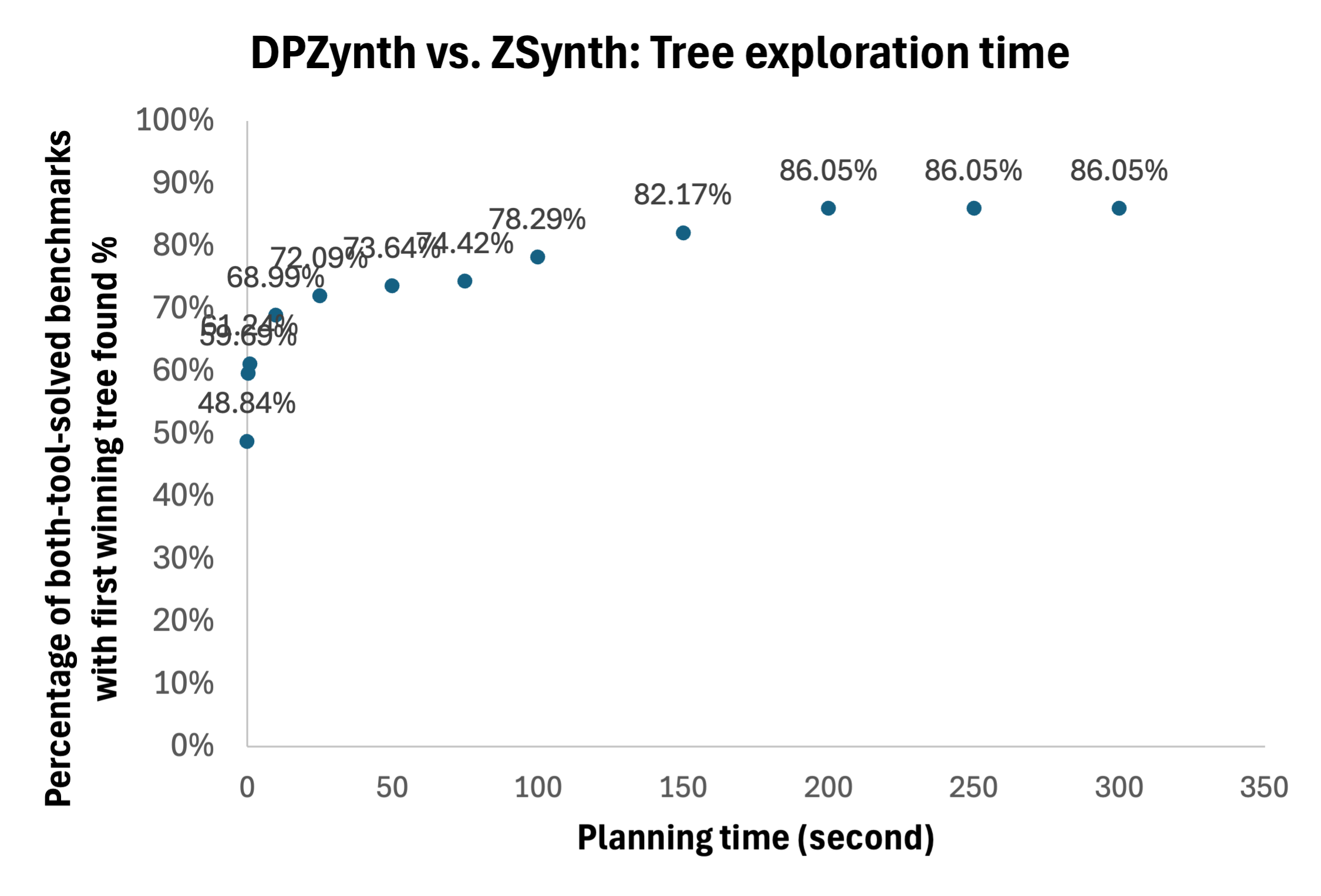}
\caption{Percentage of instances for which DPZynth is able to find a winning tree, in both-solved benchmarks, accumulated as time passes, without using magic numbers.}\label{fig: percentage at least one}
\end{subfigure}
\hfill
\begin{subfigure}
\centering
\includegraphics[width=\textwidth]{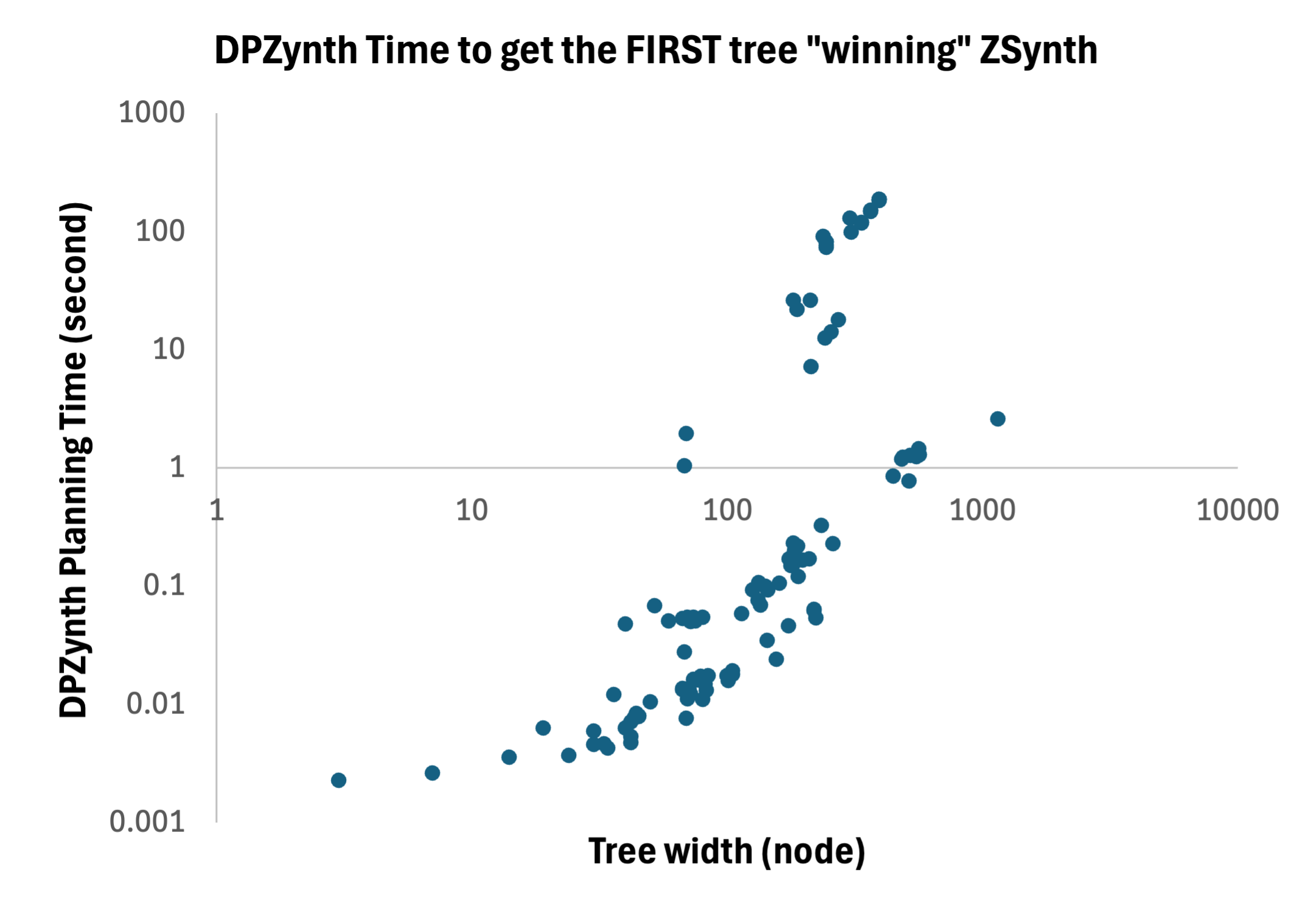}
\caption{Planning time to find a winning tree}\label{fig: first tree time}
\end{subfigure}
\caption{}
\end{figure}

To choose the appropriate time-out limit, we conducted the following experiment. We set $0.1$, $0.5$, $1$, $10$, $25$, $50$, $75$, $100$, $150$, $200$, $250$ and $300$
seconds as time bounds for the tree-decomposition phase to see for how many benchmarks, under each time bound, the DPZynth tool is able to win over the monolithic tool ZSynth with respect to \emph{end-to-end} realizability-checking time. 
This comparison is depicted in Figure~\ref{fig: percentage at least one}, which shows the percentage of benchmarks with winning trees, out of the $129$ benchmarks that both DPZynth and ZSynth could solve for realizability. After $200$ seconds there is no significant improvement, which is why we chose 200 as the default time out for tree decomposition.

To identify the target tree width, we ran a second experiment.
For each benchmark, we used the tree-decomposition tool to find \emph{all} trees within $30$-minute planning time. Then we ran the execution phase on each tree for realizability checking. We then computed the \emph{end-to-end} time for each tree, and identified, for each benchmark, the \emph{first} tree that is a \emph{winning} tree as defined above, i.e., where end-to-end time for DPZynth is shorter than that of ZSynth. (Note that there are some benchmarks where no winning trees were found within $30$-minute planning time.) Figure~\ref{fig: first tree time} depicts the planning time measured to find the first winning tree. As tree width increases, the time needed to find the first \emph{winning} tree 
is usually shorter than $200$ seconds. 
Furthermore, the $3/4$ quartile of tree widths of the first winning trees is $219$ for the benchmarks that DPZynth wins over ZSynth, and both tools can solve realizability checking, and the median is $83.5$, so we chose $200$ as the magic number for target tree width. An analogous evaluation, but considering also witness generation, confirmed this decision.


\subsection{Realizability Checking}\label{subsec: experiments realizability part}

\begin{figure}[h]\centering
\centering
\includegraphics[width=0.5\textwidth]{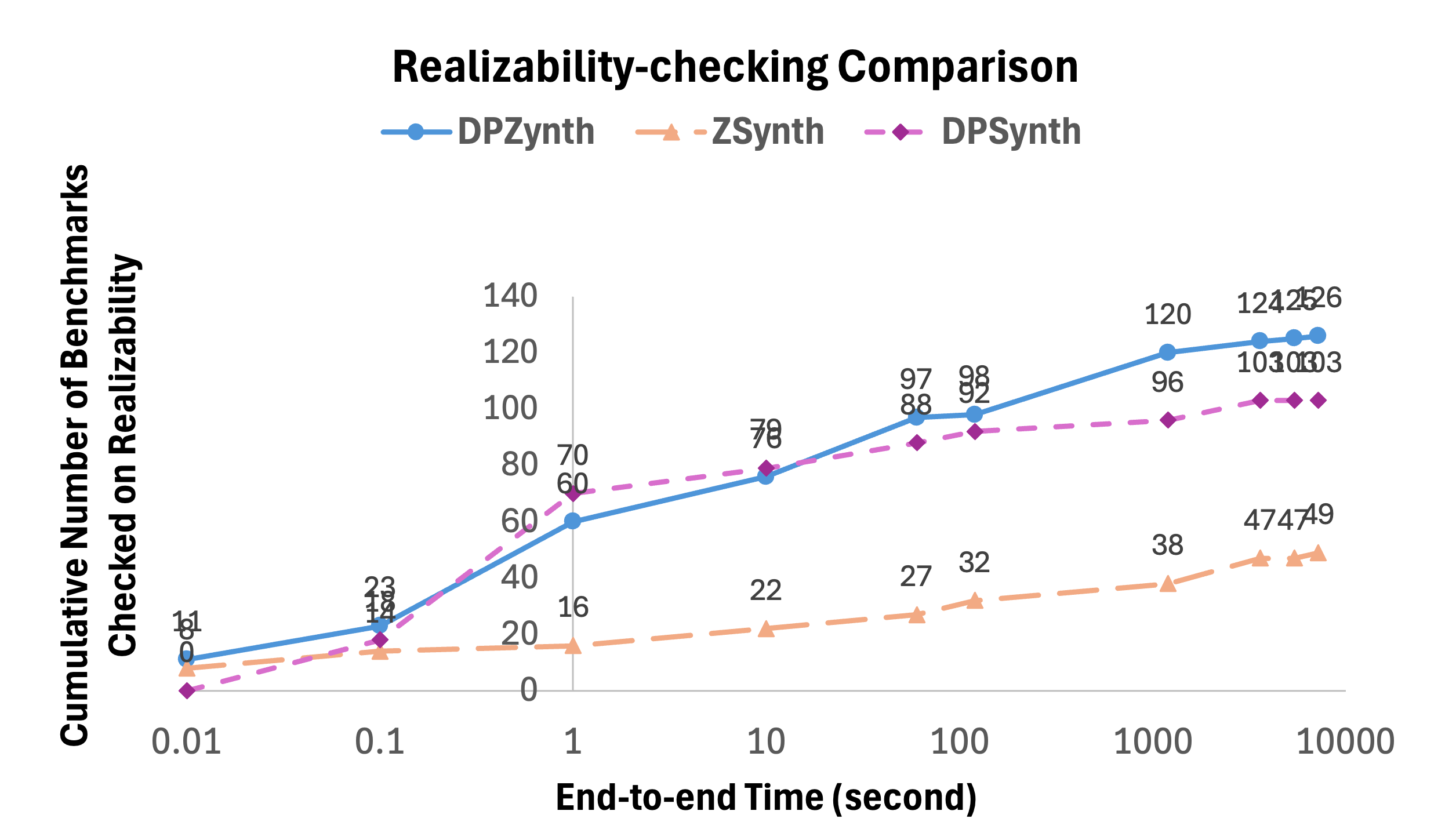}
\caption{Comparison on End-to-end Realizability Time}\label{fig: realizability time chart}
\end{figure}

Accordingly, we configured DPZynth with 200 as the magic number for both planning-timeout limit and treewidth target. First, run the conversion from a CNF formula to ZDD. Then, set $200$ seconds as planning time bound and let the tree-decomposition tool search for graded project-join trees of width $<=200$. If such a tree is found in less than $200$ seconds, we terminate planning and use that tree in the execution phase. If no such tree is found within $200$ seconds, we select the tree with the smallest tree width found within $200$ seconds and proceed to the execution phase checking realizability, with a time limit of two hours. In the case that the tree decomposition tool is unable to find any graded project-join tree within $200$ seconds, we just let the tool stop there and treat it as DPZynth is not able to solve for that benchmark. 


The results of the realizability-checking experiment are in Figure~\ref{fig: realizability time chart}. 
The overhead of the planning phase does affect the performance at the start, but time passes the advantage of the dynamic-programming approach increases. Similarly, the advantage of using ZDDs over BDDs increases over time. 
The relative performance on space consumption is similar to those on time.

\subsection{Synthesis}

\begin{figure}[h]\centering
\centering
\includegraphics[width=0.5\textwidth]{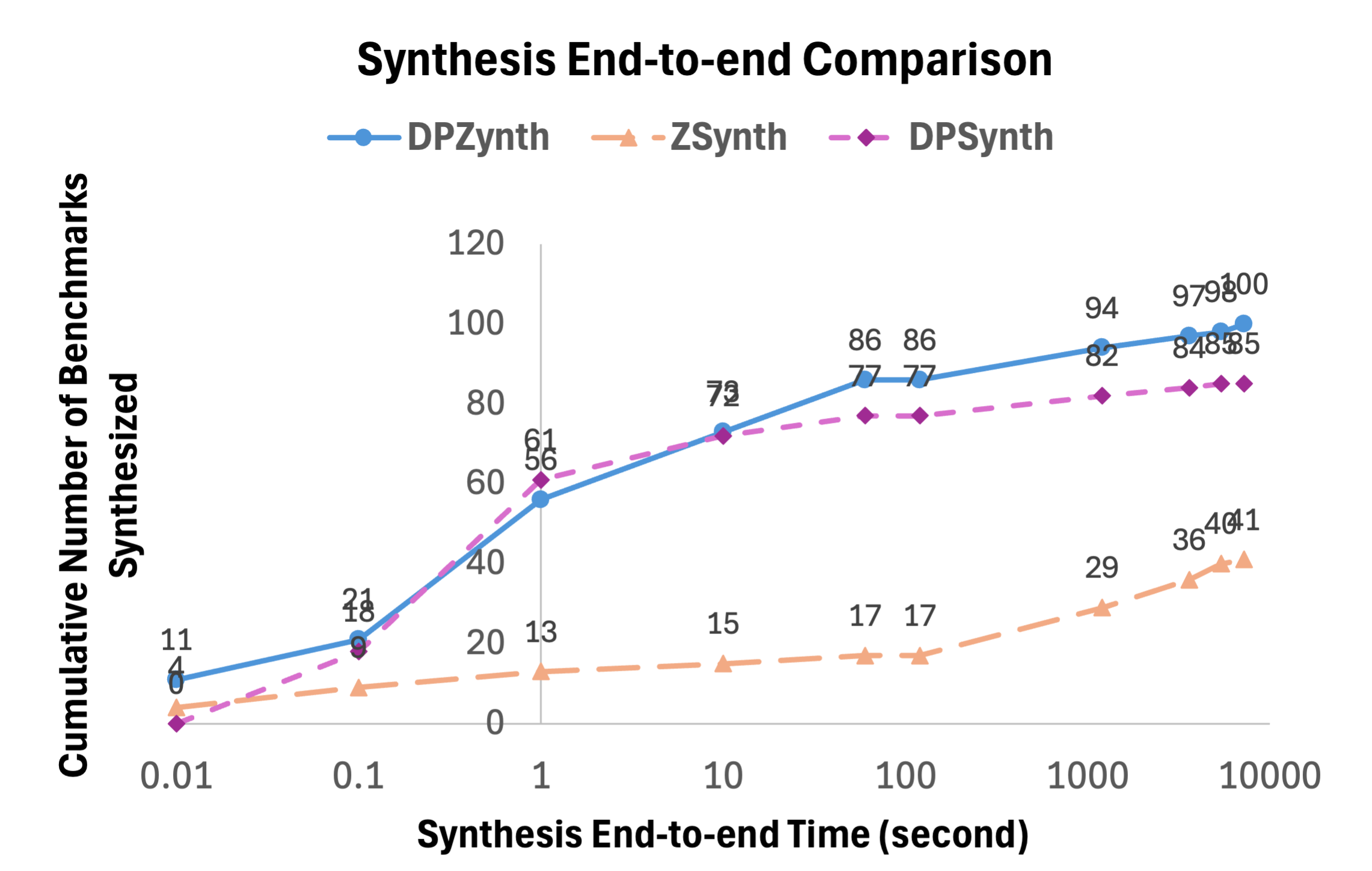}
\caption{Comparison on End-to-end Synthesis Time}\label{fig: synth time chart}
\end{figure}

Using $200$ as the magic number for end-to-end comparison up to two-hour time bound, where \emph{end-to-end} here includes also witness construction time, we see in Figure~\ref{fig: synth time chart} a similar phenomenon to realizability-checking comparison.
The speedup related to dynamic programming is still there, as well as the advantage of ZDDs compared to BDDs as time passes, although a slightly reduced advantage of ZDD is seen. 

\begin{figure}[h]\centering
\begin{subfigure}{6.5cm}
\centering
\includegraphics[width=\textwidth]{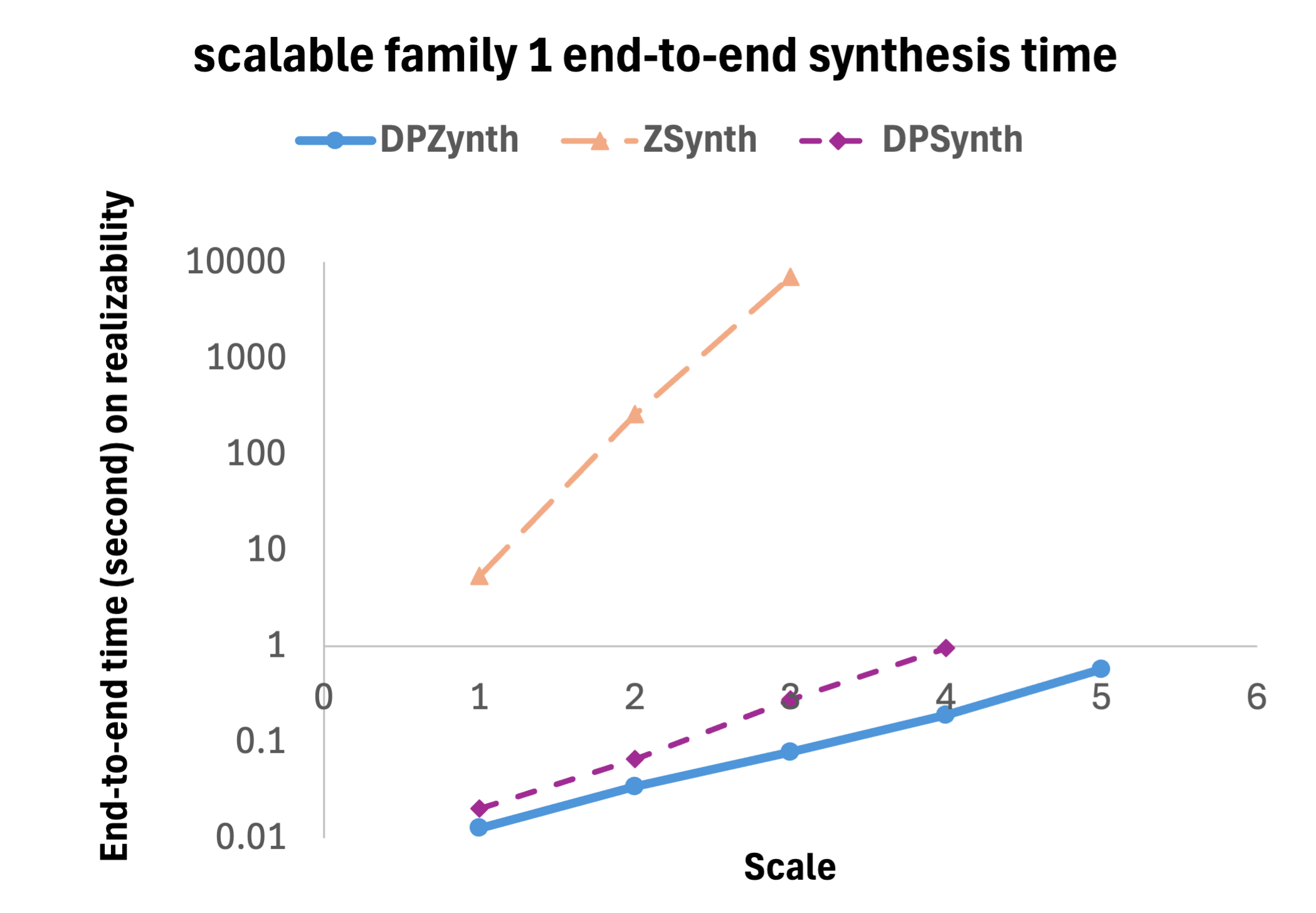}
\caption{} \label{fig: synth scalable family 2a}
\end{subfigure}
\hfill
\begin{subfigure}{6.5cm}
\centering
\includegraphics[width=\textwidth]{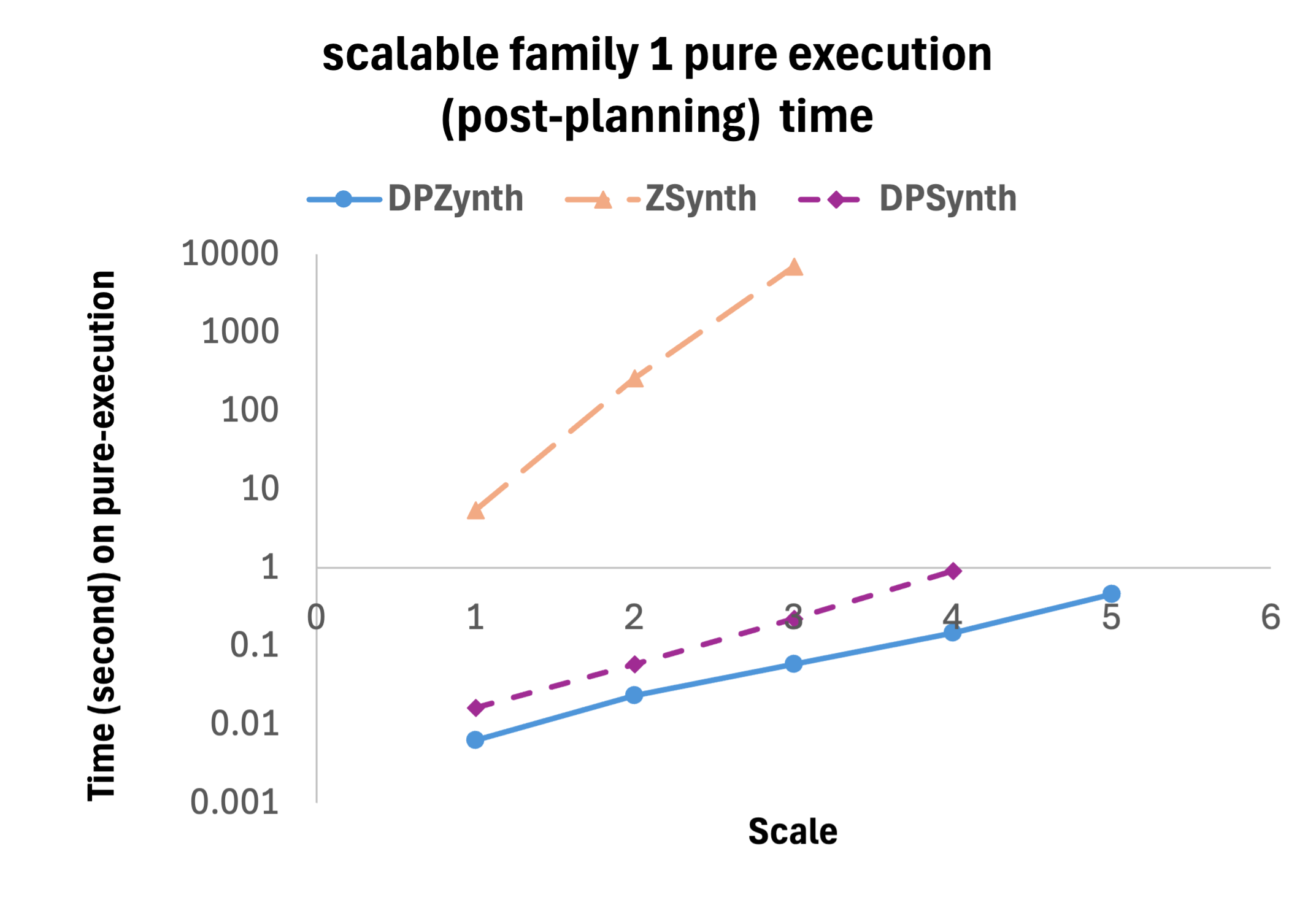}
\caption{} \label{fig: synth scalable family 2b}
\end{subfigure}
\vfill
\begin{subfigure}{6.5cm}
\centering
\includegraphics[width=\textwidth]{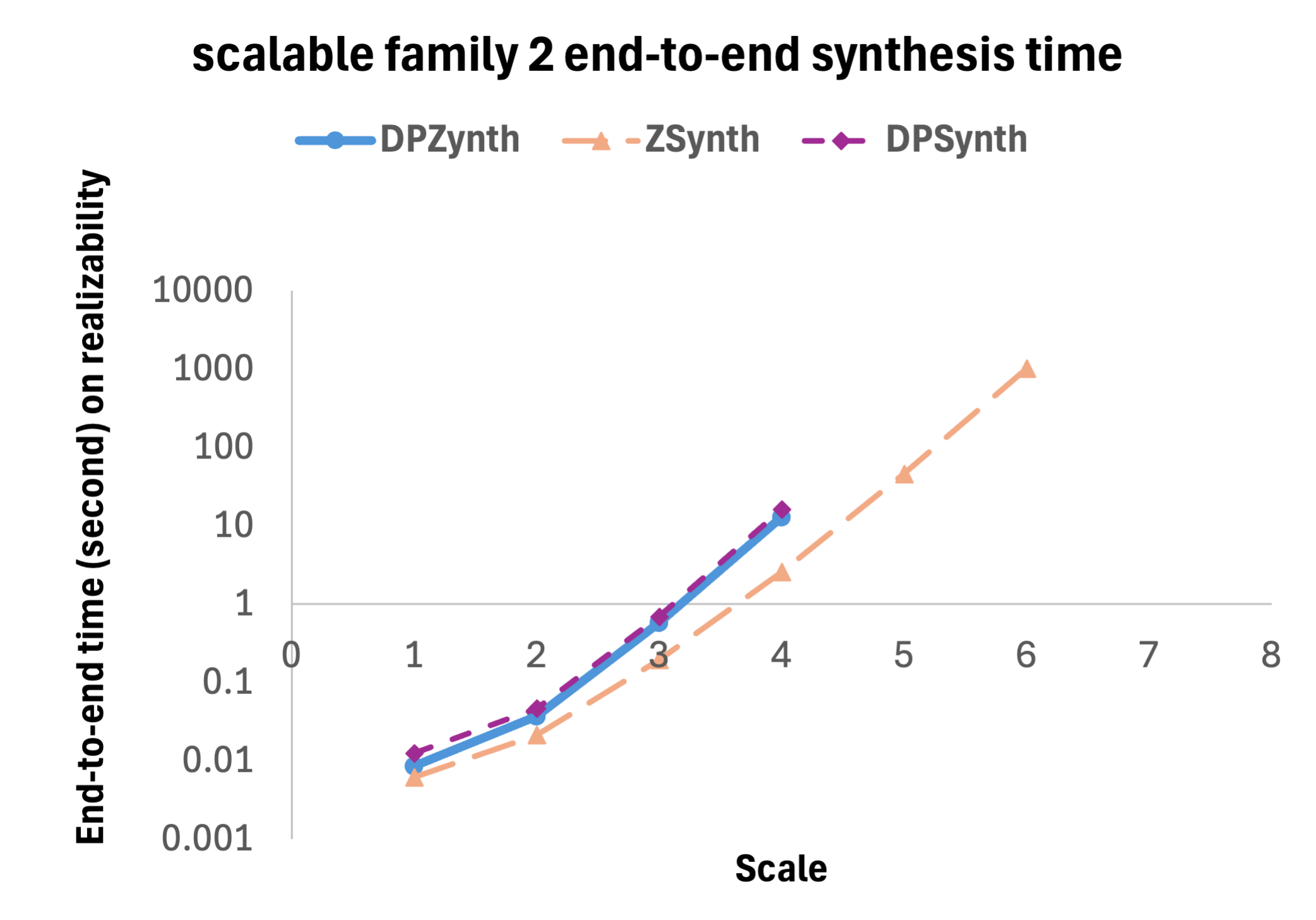}
\caption{}\label{fig: synth scalable family 1a}
\end{subfigure}
\hfill
\begin{subfigure}{6.5cm}
\centering
\includegraphics[width=\textwidth]{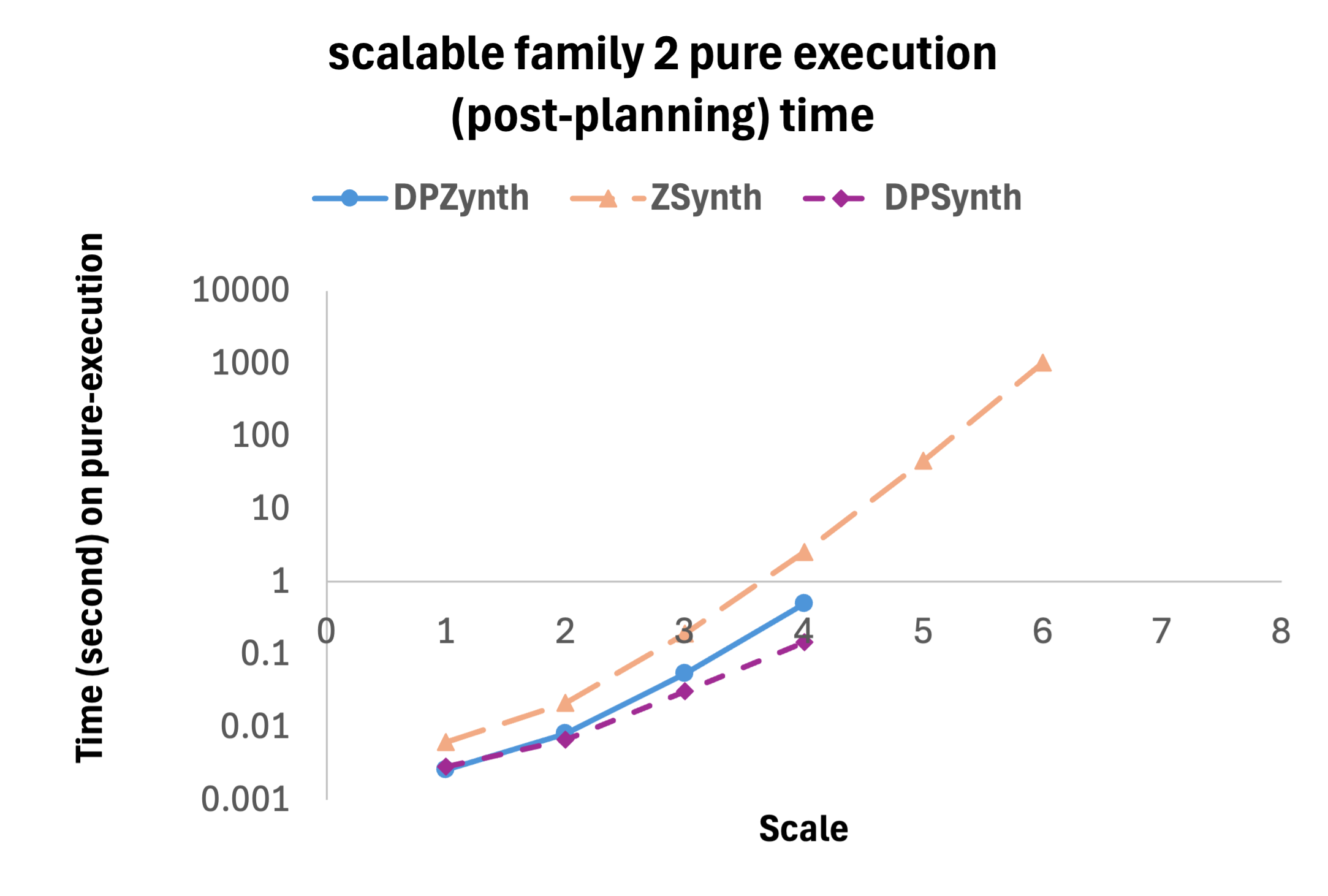}
\caption{}\label{fig: synth scalable family 1b}
\end{subfigure}
\caption{}\label{fig: synth scalable family 1}
\end{figure}

A somewhat different picture emerges when we consider scalable families of benchmarks \cite{charwat2016bdd}.
In the \texttt{mutex} family, DPZynth performs exponentially better than ZSynth and DPSynth; see
Figures~\ref{fig: synth scalable family 2a} and Figure~\ref{fig: synth scalable family 2b}. Furthermore, DPSynth and DPZynth were able to solve larger problems that ZSynth.
For benchmarks in the \texttt{qshifter} family, however, planning time becomes a dominant factor and ZSynth outperforms both DPSynth and DPZynth;
see Figures~\ref{fig: synth scalable family 1a} and~\ref{fig: synth scalable family 1b}. Thus, the magic number 200 is not a ``magic solution''.  While it leads to superior performance across our full benchmark suite, it may not perform so well on important subsets of the full suite.



The exploration-exploitation trade-off is called a ``dilemma'' because it defies a formulaic solution. Deciding how much time to explore in order to improve exploitation time is a challenging problem, whose solution is clearly workload dependent.  As we discuss in the Conclusions, we believe that machine learning offers a way to address this dilemma.


%% file: Conclusion.tex
\section{Conclusions}\label{sec: conclusion}
In this work we proposed an algorithm using a dynamic-programming framework and ZDD representation for symbolic boolean synthesis. The algorithm first checks realizability and identifies realizable instances, and then synthesizes the witness functions. We adapted monolithic realizability checking and synthesis algorithms to the dynamic-programming framework.  By evaluating the performance of the accompanied synthesis tool DPZynth, we investigated the exploration-and-exploitation dilemma exemplified by balancing planning time in tree decomposition and execution time for actual solving. We resolved this dilemma by proposing a ``magic number'' methodology that can also be used in industrial settings with a tree-decomposition time limit and tree width upper bound. As most industrial solvers are multi-engine~\cite{cabodi2013thread,multi-engine}, we believe that dynamic-programming ZDD-based tools ought to be included in the portfolio of boolean-synthesis solvers, just as previous works have emphasized the value of BDD-based synthesizers.

For future work, a major and the most important research idea to explore is using machine-learning techniques to identify the best engine for a specific instance~\cite{satzilla} or to select a right bound for the magic number for solver configuration for a problem instance \cite{hoos2012programming}.
Machine learning often requires a large dataset, so such an investigation would likely require access to industrial benchmark suites.

There are also other relevant directions to explore. In terms of optimizing the ZDD structure, low-level decision-diagram operations and variable orderings~\cite{rudell1993dynamic} are open to future research. In terms of applications to a wider scope of problems, the applicability of dynamic-programming ZDD-based framework to symbolic model checking~\cite{burch1992symbolic}, temporal synthesis~\cite{ZhuTLPV17}, and construction of unrealizablity certificates of CNFs~\cite{bryant2021generating,bryant2021dual} are possible future exploration topics.

%% file: Appendix.tex
\newpage

\appendix
\section{Proofs} \label{sec: appendix}

\subsection{Proof of Lemma.~\ref{lemma: resolution}}
\noindent
\input{proofs/proof_lemma1}

\subsection{Proof of Theorem.~\ref{thm: alg computeValuation correctness}}
\noindent
\input{proofs/proof_thm1}

\subsection{Proof of Theorem.~\ref{thm: alg RealizeZR correctness}}
\noindent
\input{proofs/proof_thm2}

\subsection{Proof of Theorem.~\ref{thm: alg compute zdd of R correctness}}
\noindent
\input{proofs/proof_thm3}

\subsection{Proof of Theorem.~\ref{thm: synth alg}}
\noindent
\input{proofs/proof_alg4}

%% file: proofs/proof_lemma1.tex
\begin{proof}
The claim follows from \cite{ZDDpaper}, where it is proved that the boolean formula ${\Sigma}_y{\varphi} $ is logically equivalent to $(( \varphi^+_y \lor \varphi^-_y)\land \varphi'_y)$.
\end{proof}

%% file: proofs/proof_thm1.tex
\begin{proof}
    We prove by structural induction on descendant levels of nodes. 
    
    In the base case where $n$ is a leaf, the sub-tree $\T_n$ has only one level, line~\ref{line leaf} is obviously equivalent to definitions of $\texttt{BV}_{\texttt{post}}(\T, n)$ and $\texttt{BV}_{\texttt{pre}}(\T, n)$ in Definition~\ref{def: ZDD pair valuations}.

    As inductive step, assume $n$ is a node in $\T_n$, and all descendants of $n$ have been computed with their correct pre and post valuations in the recursive call in line~\ref{line recursive call}. We need to show either the valuations returned in line~\ref{line return} are correct for $n$, or the formula has nullary realizability and the algorithm returns in lines~\ref{line return 1}, or \ref{line return 3}.

    The pair of valuations for $m \in C(n)$ is computed in its line~\ref{line recursive call} of the loop of its parent. If Algorithm~\ref{alg: compute valuations} returns in line~\ref{line return}, the following conditions are ensured to hold true:

    \begin{itemize}
        \item for all $m \in C(n)$, $\texttt{BV}_{\texttt{post}}(\T_m, m) \equiv \texttt{BV}_{\texttt{post}}(\T_n, m)$ is subsumption-freely unioned with $\texttt{BV}_{\texttt{pre}}(\T_n, n)$, i.e., the set of models of the formula represented by $\texttt{BV}_{\texttt{pre}}(\T_n, n)$ is a subset of the models of the formula encoded by $\texttt{BV}_{\texttt{post}}(\T_m, m)$ for each $m \in C(n)$
        \item for all $v \in \pi\mid_m(m)$, we have that $v \not\in \pi\mid_n(n)$,
        \item The base for subsumption-free unions is the empty set.
    \end{itemize}
    These invariants guarantee that the returned valuations are consistent with Definition~\ref{def: ZDD pair valuations}.

    Now we are left with proving the early returned nullary realizability in lines~\ref{line return 1} or \ref{line return 3} are correct. As early returns only occurs when either $\texttt{post-BV}(\T_n, n) == \{\emptyset\}$, $\texttt{pre-BV}(\T_n, n) == \{\emptyset\}$, or $\texttt{post-BV}(\T_m, m) == \{\emptyset\}$ for some $m \in C(n)$ (which still belongs to the case that $\texttt{pre-BV}(\T_n, n) == \{\emptyset\}$. Hence when nullary is returned, the algorithm also has the correct result.

\end{proof}

%% file: proofs/proof_thm2.tex
\begin{proof}

    By Definition~\ref{def: ZDD pair valuations}, 
    $\texttt{BV}_{\texttt{post}}(\T, n)
    = 
    {\Sigma}_{\pi(n)}( \texttt{BV}_{\texttt{pre}}(\T, m_1) \bigcup_{sf} \ldots \bigcup_{sf} \texttt{BV}_{\texttt{pre}}(\T, m_k) )$, where $C(n)=\{m_1, \ldots, m_k\}.$ Therefore, for an $n \in \texttt{YTreeRoots}(\T)$, $\texttt{BV}_{\texttt{post}}(\T, n)={\Sigma}_{\pi(n)}( \texttt{BV}_{\texttt{pre}}(\T_{m_1}, m_1) \bigcup_{sf} \ldots \bigcup_{sf} \texttt{BV}_{\texttt{pre}}(\T_{m_k}, m_k) )$. The subsumption-free union of all ZDDs $\texttt{BV}_{\texttt{post}}(\T, n)$ for all $n \in \texttt{YTreeRoots}(\T)$ is computed by line~\ref{line union} of Algorithm~\ref{alg: compute zr}, which is equivalent to taking the union of ZDDs representing clauses in all leaves $l \in \Lv{\T} \bigcap D(n)$, then projecting out all variables $v \in \pi\mid_n$ for all $n \in \texttt{YTreeRoots}(\T)$.
    Then it follows from Definition~\ref{def: full partial null real} that:
    Throughout the updates in Algorithm~\ref{alg: compute zr} lines~\ref{line update 1} and~\ref{line update 2}, the following invariants are persist to hold:
    \begin{itemize}
        \item $\bigcup_{n \in \V{\T_{\texttt{new}}}} \pi_{\texttt{new}}(n)$ contains all variables in the formula represented by $\T_{\texttt{new}}$ in the current iteration,
        \item In an iteration the current $\texttt{BV}_{\texttt{post}}(T_{\texttt{new}}, l_1) \bigcup_{sf} \ldots \bigcup_{sf} \texttt{BV}_{\texttt{post}}(T_{\texttt{new}}, l_m) )$, where $\Lv{\T_{\texttt{new}}}=\{l_1, \ldots, l_m\}$, is always equivalent to the current union $\texttt{BV}_{\texttt{post}}(\T, l'_1) \\ \bigcup_{sf} \ldots \bigcup_{sf} \texttt{BV}_{\texttt{post}}(\T, l'_k) )$, where $\texttt{XTreeLeaves}(\T)=\{l'_1, \ldots, l'_k\}$.
    \end{itemize}

\end{proof}

%% file: proofs/proof_thm3.tex
\begin{proof}
    Given Theorem~\ref{thm: alg RealizeZR correctness}, $\llbracket R \rrbracket$ at the end of Algorithm~\ref{alg: compute zr} is a representation of the realizability set. Next we show when Algorithm~\ref{alg: realize zr} returns fully realizable, $\llbracket R \rrbracket = \{\}$. This case occurs in line~\ref{line {} alg 3} of Algorithm~\ref{alg: realize zr}, when the returned $\llbracket R \rrbracket$ with respect to $\T_{\texttt{new}}$ is $\{\}$. 
    
    On the other hand, the only case when Algorithm~\ref{alg: realize zr} returns nullary realizable is $\AlgOne(\T_{\texttt{new}})$ returns $\post(\T_{\texttt{new}}, r) == \{\emptyset\}$, which only includes the case that of early determination in Algorithm~\ref{alg: compute valuations} and~\ref{alg: compute zr} identifies either (1) on the case where the result of subsumption-free union on post valuations of some nodes in $\texttt{YTreeRoots}(\T)$ or their descendants are already boolean contradiction, hence there is no need to apply additional unions given the fact that no solutions exist, or (2) on the case when projections on some variables made the solution set no longer possible to be satisfiable. Hence the rest of the cases are partially realizable.
\end{proof}

%% file: proofs/proof_alg4.tex
\begin{proof}
For variables in $\I_Y$, they are either in $\texttt{YTreeRoots}(\T)$ or the descendants of some other nodes in $\texttt{YTreeRoots}(\T)$. The order that they are added into $V$ before line~\ref{line: got V} ensures that ancestors are always added before descendants. By line~\ref{line: got V}, $V$ is the set of all nodes in $\I_Y$. By~\cite{ZDDpaper}, we know $g_{i,CNF}$ obtained in line~\ref{line on-node labeled vars} are correct functions such that $\texttt{BV}_\texttt{pre}(\T, n)\forall_{i \in \pi(n)}[y_i \mapsto g_{i,CNF}][\neg y_i \mapsto g_{i,DNF}]=1$ for $n \in V$. Note that we distinguish positive and negative literals in the substitution operation here, redefining the usual semantical interpretation of the $\mapsto$ symbol. As the second invariant, line~\ref{line A(n) substitution} preserves this property, i.e., $\texttt{BV}_\texttt{pre}(\T, n)\forall_{i \in \pi(n)}[y_i \mapsto g_{i,CNF}][\neg y_i \mapsto g_{i,DNF}]\forall_{j \in \bigcup_{n \in D(n')} \pi(n')}[y_j \mapsto g_{j,CNF}][\neg y_j \mapsto g_{j,DNF}]=1$ for $n \in \I_Y$ as $g_{i,CNF}$ for $i \in \pi(n)$ are formulas of variables including $\vec{x}$ and $j \in \bigcup_{n \in D(n')} \pi(n')$. Therefore, we know first the witnesses constructed in iteration of node $n \in V$ in the for loop in lines~\ref{line for loop start} to~\ref{line for loop ends} are witnesses to the formula represented by $\texttt{BV}_\texttt{pre}(\T, n)$. The parent node of $n \in \Lv{\T}$ is in $\I_Y$, as we assume no purely-$X$ clauses. 

Next, we show by induction that for all nodes $n \in \I_Y\cup \Lv{\T}$, if we substitute the pairs of witnesses into literals of all output variables labeled on its ancestors up to $\texttt{YTreeRoots}(\T)$, the pre-valuation of the node is 1. That is, 
\begin{align*}
    \texttt{BV}_\texttt{pre}(\T, n)\forall_{y_i \in \cup_{n \in D(m), m \in \I_Y)} \pi(m)}y_i [y_i \mapsto g_{i,CNF}][\neg y_i \mapsto g_{i,DNF}]=1. \tag{1}
\end{align*} is true for all $n \in \I_Y\cup \Lv{\T}$.

In the base case, when $n \in \texttt{YTreeRoots}(\T)$, 
it is shown above that (1) is true.

For inductive step, if (1) is true for parent node $n'$ of $n$, then  

\begin{align*}
    &\texttt{BV}_\texttt{pre}(\T, n')\forall_{y_i \in \cup_{n' \in D(m), m \in \I_Y} \pi(m)}y_i [y_i \mapsto g_{i,CNF}][\neg y_i \mapsto g_{i,DNF}]=1\\
    &={\bigcup_{sf}}_{n \in C(n')}\texttt{BV}_\texttt{post}(\T, n)\forall_{y_i \in \cup_{n' \in D(m), m \in \I_Y} \pi(m)}y_i [y_i \mapsto g_{i,CNF}][\neg y_i \mapsto g_{i,DNF}] \tag{where ${\bigcup_{sf}}_{n \in C(n')}\texttt{BV}_\texttt{post}(\T, n)$ is equivalent to encoding of the }\\\tag{conjunctions of formulas represented by $\texttt{BV}_\texttt{post}(\T, n).$}\\
\end{align*}

Hence $\texttt{BV}_\texttt{post}(\T, n)\forall_{y_i \in \cup_{n' \in D(m), m \in \I_Y} \pi(m)}y_i [y_i \mapsto g_{i,CNF}][\neg y_i \mapsto g_{i,DNF}] = 1$ for $n\in C(n')$. But it is also equivalent to $(\Sigma \pi(n))\texttt{BV}_\texttt{pre}(\T, n)\forall_{y_i \in \cup_{n \in D(m), m \in \I_Y} \pi(m)}\\y_i [y_i \mapsto g_{i,CNF}][\neg y_i \mapsto g_{i,DNF}]$.

So $\texttt{BV}_\texttt{pre}(\T, n) \forall_{y_i \in \cup_{n \in D(m), m \in \I_Y} \pi(m)}y_i [y_i \mapsto g_{i,CNF}][\neg y_i \mapsto g_{i,DNF}]=1$ for $\forall n \in C(n')$.

Therefore, (1) holds true for all $n \in \I_Y\cup \Lv{\T}$.

We also have $R = \llbracket (\Sigma_{y \in \vec{y}}) \varphi \rrbracket= \bigcup_{sf} \texttt{BV}_\texttt{post}(\T, n)$ of all $n \in \texttt{YTreeRoots}(\T)$ by Theorem~\ref{thm: alg RealizeZR correctness} in Section~\ref{sec:realizability}, which encodes the conjunction of formulas represented by $\texttt{BV}_\texttt{post}(\T, n)$ of all $n \in \texttt{YTreeRoots}(\T)$. 

On the other hand, we know that $\texttt{BV}_\texttt{post}(\T, n)$ for $n \not\in \Lv{\T}$ is equivalent to $\Sigma_{\pi(n)}\texttt{BV}_\texttt{pre}(\T, n)$ by Definition~\ref{def: ZDD pair valuations}. 

Thus \begin{align*}
    R = \llbracket (\Sigma_{y \in \vec{y}}) \varphi \rrbracket
    &= {\bigcup_{sf}}_{n \in \texttt{YTreeRoots}(\T)} \texttt{BV}_\texttt{post}(\T, n)\\
    &={\bigcup_{sf}}_{n \in \texttt{YTreeRoots}(\T)} \Sigma_{\pi(n)}\texttt{BV}_\texttt{pre}(\T, n).  \tag{2}
\end{align*}

We next use induction to show that the pre-valuation is equivalent to the subsumption-free union of ZDDs of leaves that are its descendants, with all variants labeled on its descendants projected, i.e., \begin{align*}
    \texttt{BV}_\texttt{pre}(\T, n) &= (\Sigma_{\pi(n')}) {\bigcup_{sf}}_{n' \in D(n)\cap\Lv{\T} }\llbracket\gamma(n')\rrbracket.\tag{3}
\end{align*}

In the base case, all children of $n$ are leaves nodes. \begin{align*}
\texttt{BV}_\texttt{pre}(\T, n) 
&={\bigcup_{sf}}_{n' \in C(n)}\llbracket\gamma(n')\rrbracket\\
&=(\Sigma \pi(n')) {\bigcup_{sf}}_{n' \in D(n)\cap\Lv{\T} }\llbracket\gamma(n')\rrbracket.
\end{align*}

For inductive step, the inductive hypothesis is that for all $n' \in C(n)$, 
\begin{align*}
    \texttt{BV}_\texttt{pre}(\T, n') &= (\Sigma_\pi(n'')) {\bigcup_{sf}}_{n'' \in D(n')\cap\Lv{\T} }\llbracket\gamma(n'')\rrbracket.
\end{align*} 

Then for $n \in \I_Y=V$, 
\begin{align*}
    \texttt{BV}_\texttt{pre}(\T, n) 
&={\bigcup_{sf}}_{n' \in C(n)}\texttt{BV}_\texttt{post}(\T, n') \\
&={\bigcup_{sf}}_{n' \in C(n)}(\Sigma \pi(n'))\texttt{BV}_\texttt{pre}(\T, n') \\
&={\bigcup_{sf}}_{n' \in C(n)}(\Sigma \pi(n'))(\Sigma \pi(n'')) {\bigcup_{sf}}_{n'' \in D(n')\cap\Lv{\T} }\llbracket\gamma(n'')\rrbracket\\
&= (\Sigma \pi(n')) {\bigcup_{sf}}_{n' \in D(n)\cap\Lv{\T} }\llbracket\gamma(n')\rrbracket. 
\end{align*}

Therefore, (3) is true for all $n \in \I_Y=V$.

Hence \begin{align*}
    R &= \llbracket (\Sigma_{y \in \vec{y}}) \varphi \rrbracket\\
    &= {\bigcup_{sf}}_{n \in \texttt{YTreeRoots}(\T)} \texttt{BV}_\texttt{post}(\T, n)\\
    &={\bigcup_{sf}}_{n \in \texttt{YTreeRoots}(\T)} \Sigma_{\pi(n)}\texttt{BV}_\texttt{pre}(\T, n)\\
    &={\bigcup_{sf}}_{n \in \texttt{YTreeRoots}(\T)} \Sigma_{\pi(n)}(\Sigma_{\pi(n')}) {\bigcup_{sf}}_{n' \in D(n)\cap\Lv{\T} }\llbracket\gamma(n')\rrbracket\\
    &={\bigcup_{sf}}_{n \in \texttt{YTreeRoots}(\T)} \Sigma_{\pi(n)} {\bigcup_{sf}}_{n' \in \Lv{\T}\cap D(n) }\llbracket\gamma(n')\rrbracket\\
    &={\Sigma_{\cup_{n \in \texttt{YTreeRoots}(\T)}}}_{\pi(n)} {\bigcup_{sf}}_{n \in \texttt{YTreeRoots}(\T)}  {\bigcup_{sf}}_{n' \in \Lv{\T}\cap D(n) }\llbracket\gamma(n')\rrbracket\tag{as variables only occur on descendants of their labels}\\
    &={\Sigma_{y \in \vec{y}}} {\bigcup_{sf}}_{n \in \texttt{YTreeRoots}(\T)}  {\bigcup_{sf}}_{n' \in \Lv{\T}\cap D(n) }\llbracket\gamma(n')\rrbracket\\
    &={\Sigma_{y \in \vec{y}}} {\bigcup_{sf}}_{n' \in \Lv{\T} }  \llbracket\gamma(n')\rrbracket. \tag{since we assume no pure-$X$ clauses}\\
\end{align*}

As we have known from (2) that $R = \llbracket (\Sigma_{y \in \vec{y}}) \varphi \rrbracket = (\Sigma_{y \in \vec{y}}) \llbracket \varphi \rrbracket$. Thus $\llbracket \varphi \rrbracket = {\bigcup_{sf}}_{n' \in \Lv{\T} } \llbracket\gamma(n')\rrbracket$. 

Then 
\begin{align*}
    &\llbracket  \varphi\rrbracket {\forall_{y_i}} [y_i \mapsto g_{i, CNF}][\neg y_i \mapsto g_{i, DNF}] \\
    =& ({\bigcup_{sf}}_{n' \in \Lv{\T} } \llbracket\gamma(n')\rrbracket) [y_i \mapsto g_{i, CNF}][\neg y_i \mapsto g_{i, DNF}]\\
    =&( {\bigcup_{sf}}_{n' \in \Lv{\T} } \texttt{BV}_\texttt{pre}(\T, n') )    [y_i \mapsto g_{i, CNF}][\neg y_i \mapsto g_{i, DNF}]\\
    =&{\bigcup_{sf}}_{n' \in \Lv{\T} } (\texttt{BV}_\texttt{pre}(\T, n')     [y_i \mapsto g_{i, CNF}][\neg y_i \mapsto g_{i, DNF}])\\
    =&1 \tag{by (1)}.
\end{align*}

Therefore, the functions constructed in Algorithm~\ref{alg: synth} are witnesses to the original CNF $\varphi$.

\end{proof}